# Evidence for microscopic kurtosis in neural tissue revealed by Correlation Tensor MRI

Rafael Neto Henriques[1], Sune N Jespersen[2,3] and Noam Shemesh[1*]

[1]Champalimaud Research, Champalimaud Centre for the Unknown, Lisbon, Portugal
[2]Center of Functionally Integrative Neuroscience (CFIN) and MINDLab, Clinical Institute, Aarhus University, Aarhus, Denmark.
[3]Department of Physics and Astronomy, Aarhus University, Aarhus, Denmark

*Corresponding author:

Dr. Noam Shemesh, Champalimaud Research, Champalimaud Centre for the Unknown, Av. Brasilia 1400-038, Lisbon, Portugal

E-mail: noam.shemesh@neuro.fchampalimaud.org;

Phone number: +351 210 480 000 ext. #4467

# Abstract

Purpose: The impact of microscopic diffusional kurtosis (µK) –arising from restricted diffusion and/or structural disorder – remains a controversial issue in contemporary diffusion MRI (dMRI). Recently, Correlation Tensor Imaging (CTI) was introduced to disentangle the sources contributing to diffusional kurtosis, without relying on a-priori microstructural assumptions. Here, we investigated µK in *in vivo* rat brains and assessed its impact on state-of-the-art methods ignoring µK.

Methods: CTI harnesses double diffusion encoding (DDE) experiments, which were here improved for speed and minimal bias using four different sets of acquisition parameters. The robustness of CTI estimates from the improved protocol is assessed in simulations. The *in vivo* CTI acquisitions were performed in healthy rat brains using a 9.4T pre-clinical scanner equipped with a cryogenic coil, and targeted the estimation of µK, anisotropic kurtosis, and isotropic kurtosis.

Results: The improved CTI acquisition scheme substantially reduces scan time and importantly, also minimizes higher-order-term biases, thus enabling robust µK estimation, alongside $K_{aniso}$ and $K_{iso}$ metrics. Our CTI experiments revealed positive µK both in white and grey matter of the rat brain *in vivo*; µK is the dominant kurtosis source in healthy grey matter tissue. The non-negligible µK substantially biases prior state-of-the-art analyses of $K_{iso}$ and $K_{aniso}$.

Conclusion: Correlation Tensor MRI offers a more accurate and robust characterization of kurtosis sources than its predecessors. µK is non-negligible in vivo in healthy white and grey matter tissues and could be an important biomarker for future studies. Our findings thus have both theoretical and practical implications for future experiments.

# Index

# 1. Introduction

Diffusion MRI (dMRI) has become one of the most important methods for non-invasively probing microstructural features in health and in disease (1–3). Single diffusion encoding (SDE) experiments (4,5) (Fig. 1A) probing diffusion along a single axis, have been widely used to measure the directional apparent diffusion coefficient (6–8), the diffusion propagator from q-space experiments (9–12), the diffusion tensor (13–15) and its time-dependence (16–18), and the diffusional kurtosis (via Diffusional Kurtosis Imaging, (DKI), (19)). Together with microstructural and biophysical models (20–24), SDE has provided important insight into microstructural tissue.

DKI in particular provides important quantitative information on the degree of non-Gaussian diffusion reflected in the diffusional excess-kurtosis (19). DKI has been shown to be very sensitive to, inter-alia, age-related microstructural changes (25–29), ischemia (30,31), tumors (32–34), traumatic brain injury (35,36), and Parkinson's and Alzheimer's diseases (37,38). However, the biological interpretation of its measures remains limited because non-Gaussian diffusion may arise from different sources (39–41). Several compartment models have been proposed to relate non-Gaussian diffusion with their biological underpinnings (19,42–45); however, the specificity of these models can be severely compromised by their strong assumptions and constraints (46–50).

Unique information about the non-Gaussian nature of tissue diffusion can be resolved using multidimensional diffusion encoding (MDE) strategies (51–53). In contrast to SDE methods, MDE probes diffusion correlations across different dimensions by either including additional pairs of pulse gradients (54–62) or using continuous gradient waveforms with 3D trajectories (63–69). Under the strict Multiple Gaussian Components (MGC) assumption (no time-dependence and no kurtosis arising from restricted diffusion or structural disorder),

Westin et al. showed that MDE can be generally described by tensor-valued encoding information (70). Using different b-tensor shapes, MDE can resolve anisotropic and isotropic kurtosis sources ($K_{aniso}$ and $K_{iso}$), thereby reporting on the shape and size variances of components represented by diffusion tensors (39,68,71,72). Such MGC analyses are, however, doubly prone to bias arising from restricted diffusion (e.g., upon interaction with microscopic boundaries). First, the MGC analyses of MDE data biases $K_{aniso}$ and $K_{iso}$ when continuous gradient waveforms are long compared to the time it takes to probe the boundaries (67,73,74). Secondly, as we will show in this work, even for fixed diffusion times and tuned MDE, MGC analysis can be biased by non-Gaussian effects (i.e. microscopic kurtosis $\mu K$) that arises from restricted diffusion or systems presenting complex microstructural features with characteristic lengths in the order of the probed scales (40,41,49,67,73), e.g. intra-cellular cross-sectional size variance, extra-cellular tortuosity.

Double diffusion encoding (DDE, (5,54–62)) is an MDE variant probing diffusion via two pairs of diffusion pulsed gradients (Fig. 1B). Although previously used without MGC analyses to resolve microscopic anisotropy (61,75–78), recent MGC approaches harnessed DDE's inherent capability to provide both linear and planar encodings, thereby providing sufficient information to resolve $K_{aniso}$ and $K_{iso}$ based on MGC analyses. Moreover, DDE at the long mixing time regime can be used to minimize diffusion time-dependent effects (54,55,57,76,79,80). Although prior DDE studies have attempted to measure $\mu K$, orientation dispersion was inherently conflated in the method (41,81). Going beyond the MGC framework, the Correlation Tensor Imaging (CTI) approach (40) was recently introduced for $\mu K$ measurements. The CTI framework allows the simultaneous decoupling of $K_{aniso}$, $K_{iso}$ from $\mu K$ effects without resorting to the multi-Gaussian assumption (40) or without conflation with other mesoscopic effects (41). However, the initial CTI approach (40) can suffer from higher-order effects, and was quite time-consuming to acquire, thereby limiting its *in vivo* applicability (40).

Here, we aimed to investigate the existence of $\mu K$ in *in vivo* neural tissues and their impact on the increasingly popular MGC approaches. We first develop an improved and highly accelerated CTI acquisition scheme, which is more robust towards $\mu K$ estimation, with minimized high-order-term biases. We then investigate the existence of $\mu K$ in in vivo rat brains and investigate its impact on the highly popular MGC approach. Our results provide a new window for quantifying microstructure in health and disease, and show that $\mu K$ must be considered in future dMRI studies.

# 2. Theory

## 2.1. Total kurtosis estimates from single diffusion encoding

The SDE signal attenuation ($E_{SDE}$) can be expressed (with Einstein summation convention) as the following 2$^{nd}$ order cumulant expansion (19):

$$\log(E_{SDE}(b, \boldsymbol{n})) = -n_i n_j b D_{ij} + \frac{1}{6} n_i n_j n_k n_l b^2 \bar{D}^2 W_{ijkl} + O(b^3) \qquad (1)$$

where $b$ is the b-value defined by $b = (\gamma \delta g)^2 (\Delta - \delta/3)$, $\boldsymbol{n}$ is the diffusion gradient direction, $D_{ij}$ and $W_{ijkl}$ are the diffusion and excess-kurtosis tensors, and $\bar{D}$ is the mean diffusivity.

To quantify non-Gaussian diffusion decoupled from confounding effects of tissue dispersion, it is also useful to consider the cumulant expansion of powder-averaged SDE signal decays (i.e. signals averaged across multiple gradient directions) (29,48,76):

$$\log(\bar{E}_{SDE}(b)) = -b D_T + \frac{1}{6} b^2 D_T^2 K_T + O(b^3) \qquad (2)$$

where $\bar{E}_{SDE}$ is the powder-averaged SDE signal decay, $D_T$ and $K_T$ are the isotropic diffusivity ($D_T = \bar{D}$) and isotropic excess-kurtosis of powder-averaged signals ($K_T = W_{iijj}/5 + 2 D_{ij} D_{ij}/5\bar{D}^2 - 6/5$) (40). In the absence of exchange, the total kurtosis $K_T$ can be described by the sum of three different sources (40):

$$K_T = K_{aniso} + K_{iso} + \mu K \qquad (3)$$

where $K_{aniso}$ is related to tissue miscrocospic anisotropy $\mu A$ ($K_{aniso} = 2\frac{\mu A^2}{D^2}$) (39,65,76,77), and $K_{iso}$ is related to the variance of tissue components' apparent mean diffusivities $D_i$ ($K_{iso} = 3\frac{V(D_i)}{D^2}$, with $V(D_i)$ representing the variance across the mean diffusivities of tissue components) (39,51). Microscopic kurtosis $\mu K$, which was previously referred to as intra-compartmental kurtosis (40), is a weighted sum of different microscopic sources of non-Gaussian diffusion $\mu K_i$

$$\mu K = \frac{\langle D_i^2 \mu K_i \rangle}{D^2}, \tag{4}$$

with $\langle \cdot \rangle$ representing the average over tissue components). Here $\mu K_i$ can be related to non-Gaussian diffusion arising from restricted diffusion (41,82,83) or tissue disorder due to the presence of microscopic hindrances to water molecules, e.g. membranes, organelles, axonal caliber variations etc. (49,84–87). Although the total kurtosis $K_T$ can be estimated by fitting Eq. 2 to data acquired with at least two non-zero b-values, it is important to note that the kurtosis sources in Eq. 3 cannot be decoupled from SDE experiments in a model free manner.

## 2.2. Correlation Tensor Imaging kurtosis source estimation

Recently, the Correlation Tensor Imaging (CTI) methodology was proposed to resolve different kurtosis sources from DDE signals (40). Fig. 1B shows an illustration of the DDE sequence which probes diffusion using two pairs of pulsed gradients with magnitudes $g_1$ and $g_2$, widths $\delta_1$ and $\delta_2$, separations time $\Delta_1$ and $\Delta_2$, and mixing time $\tau_m$ (Fig. 1B). Note that the DDE pairs can also be applied along different directions, $\boldsymbol{n}_1$ and $\boldsymbol{n}_2$. To probe kurtosis for fixed timing parameters, CTI uses $\delta_1 = \delta_2 = \delta$ and $\Delta_1 = \Delta_2 = \Delta$. Moreover, to avoid diffusion time-dependent biases, CTI is applied to DDE data acquired at long mixing time. In this regime and up to 2$^{nd}$ order in $b$, the DDE signal attenuation ($E_{DDE}$) can be expressed as (40,60,76,80):

$$\log(E_{DDE}(b_1, b_2, \boldsymbol{n}_1, \boldsymbol{n}_2)) = -(n_{1i}n_{1j}b_1 + n_{2i}n_{2j}b_2)D_{ij}$$
$$+ \frac{1}{6}(n_{1i}n_{1j}n_{1k}n_{1l}b_1^2 + n_{2i}n_{2j}n_{2k}n_{2l}b_2^2)\overline{D}^2 W_{ijkl} + \frac{1}{4\left(\Delta-\frac{\delta}{3}\right)^2} n_{1i}n_{1j}n_{2k}n_{2l}b_1 b_2 Z_{ijkl} + O(b^3)$$

(5)

where $b_1 = (\gamma \delta g_1)^2(\Delta - \delta/3)$ and $b_2 = (\gamma \delta g_2)^2(\Delta - \delta/3)$ are the b-values associated with the two DDE gradients, and $Z_{ijkl}$ is a tensor that approaches the covariance tensor ($Z_{ijkl} \rightarrow 4C_{ijkl}\left(\Delta - \frac{\delta}{3}\right)^2$) at long mixing times. We previously showed that $K_{aniso}$, $K_{iso}$, and $\mu K$ can in theory be extracted from the tensors of Eq. 5 (40). However, our preliminary validation showed that high-order-terms $O(b^3)$ can introduce biases on the different kurtosis estimates that depend on dispersion levels. To suppress this dependence, powder-averaged DDE signals ($\bar{E}_{DDE}$) are used:

$$\log(\bar{E}_{DDE}(b_1, b_2, \theta))$$

$$= -(b_1 + b_2)\bar{D} + \frac{1}{6}(b_1^2 + b_2^2)D^2 K_T + \frac{1}{2} b_1 b_2 \cos^2\theta\, \bar{D}^2 K_{aniso}$$

$$+ \frac{1}{6} b_1 b_2 \bar{D}^2 (2K_{iso} - K_{aniso}) + O(b^3)$$

(6)

where $\theta$ is defined as the angle between the gradient directions $\mathbf{n}_1$ and $\mathbf{n}_2$. Note that several different pairs of $\mathbf{n}_1$ and $\mathbf{n}_2$ with constant $\theta$ evenly sampling a 3D unit sphere are required for powder-averaging (48,76,77) – specific pairs of gradient directions used on this study are shown below. From the parameters of Eq. 6, microscopic kurtosis can be estimated by $\mu K = K_T - K_{aniso} - K_{iso}$ (cf. Eq. 3).

## 2.3. Accelerating Correlation Tensor Imaging and increasing its robustness towards higher order effects

To accelerate the CTI acquisition, we note that only the four different sets of DDE experiments illustrated in Fig. 1C (in addition to acquisitions without diffusion sensitization, i.e. $b_1 = b_2 = 0$), are required to extract CTI's metrics. The 4 sets are as follows:

1) Powder-averaged signals with $b_1 = b_a$ and $b_2 = 0$. Note that these experiments are equivalent to SDE experiments (Fig. 1C1);

2) Powder-averaged symmetric DDE with diffusion weighting $b_1 = b_2 = b_a/2$ and parallel gradient directions ($\theta = 0°$, Fig. 1C2).

3) Powder-averaged symmetric DDE with diffusion weighting $b_1 = b_2 = b_a/2$ and perpendicular gradient directions ($\theta = 90°$, Fig. 1C3);

4) Powder-averaged symmetric and parallel DDE as 3) but with a different total b-value $b_t = b_1 + b_2 = b_b < b_a$ (Fig. 1C4). Note that all previous sets (1-3) have the same total b-value $b_t = b_1 + b_2$ equal to $b_a$.

To ensure homoscedastic $\bar{E}_{DDE}$ signals, all four experiment sets should be acquired with a similar number of gradient direction pairs for powder-averaging.

## 2.4. Specificity of the improved protocol to different kurtosis sources

The analysis to resolve different kurtosis sources proceeds as follows:

a) $\mu K$ can be extracted from the log difference of powder-averaged signals from the experiments' set #1 and #2:

b)

$$\log(\bar{E}_{DDE}(b_a, 0, 0°)) - \log\left(\bar{E}_{DDE}\left(\frac{b_a}{2}, \frac{b_a}{2}, 0°\right)\right) = \frac{1}{12} b_a^2 D^2 \mu K \tag{7}$$

b) as pointed in previous studies (e.g., (76,77,79)), $K_{aniso}$ can be extracted from the log difference of powder-averaged signal from sets #2 and #3:

$$\log\left(\bar{E}_{DDE}\left(\frac{b_a}{2}, \frac{b_a}{2}, 0°\right)\right) - \log\left(\bar{E}_{DDE}\left(\frac{b_a}{2}, \frac{b_a}{2}, 90°\right)\right) = \frac{1}{2} b_a^2 D^2 K_{aniso} \tag{8}$$

c) to decouple $\bar{D}$, $K_T$, and $K_{iso}$, powder-averaged signals also require at least two non-zero total b-values $b_t$. Therefore, DDE experiments with symmetric intensities and parallel directions for a lower total b-value $b_t = b_a$ are acquired (set #4).

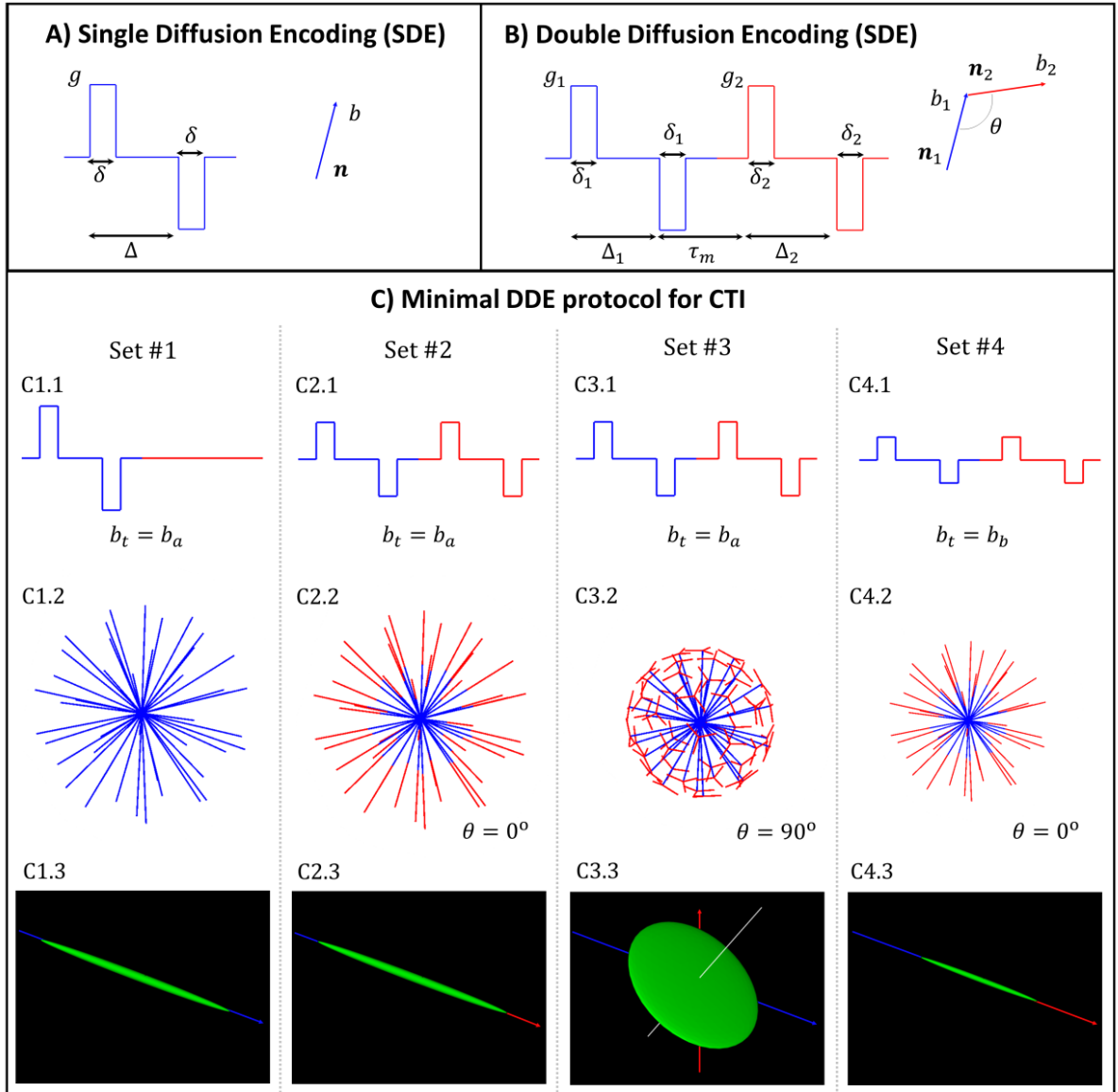

**Figure 1 – Experiments for kurtosis source estimation. A)** Parameters of a standard SDE pulse sequence, where $\Delta$ is the diffusion time, $\delta$ is the gradient pulse duration, and $g$ is the gradient pulse direction (this sequence can also be described by a gradient direction $\mathbf{n}$ and b-value $b = (\gamma g \delta)^2(\Delta - \delta/3)$); **B)** Parameters of a standard DDE pulse sequence, where $\Delta_1$ and $\Delta_2$ are the diffusion encoding blocks' diffusion times, $\delta_1$ and $\delta_2$ are their gradient pulse durations, $g_1$ and $g_2$ are their gradient intensities and $\tau_m$ is the mixing time (i.e. the time between the two diffusion encoding bocks). At the long mixing time, DDE experiments for powder-averaged systems can be fully described by two b-values ($b_1 = (\gamma g_1 \delta_1)^2(\Delta_1 - \delta_1/3)$ and $b_2 = (\gamma g_2 \delta_2)^2(\Delta_2 - \delta_2/3)$) and the angle $\theta$ between the directions of the two diffusion encoding modules $\mathbf{n}_1$ (blue) and $\mathbf{n}_2$ (red); **C)** Parameters for the different data sets required to estimate different kurtosis sources. The diffusion encoding profiles for all data sets are shown in panels C1.1, C2.1, C3.1, and C4.1. The gradient directions used for signal powder-average calculation are shown in panels C1.2, C2.2, C3.2, and C4.2. The equivalent b-tensor shapes for each experiment are shown in panels C1.3, C2.3, C3.3, C4.3.

## 2.4. Diffusion tensor variance approach

Under the MGC assumption, previous studies showed that $K_{aniso}$ and $K_{iso}$ can be estimated from signals measured using any MDE sequence that probes different b-tensor magnitudes $b_t$ and shapes. For the sake of simplicity, here we only consider axial tensor-valued experiments (39,51,72), where the b-tensor shape is characterized by a single parameter $b_\Delta \in [-\frac{1}{2}, 1]$. Thus, the powder-averaged signal is

$$\log(\bar{E}_{MGC}(b, b_\Delta)) = -b_t \bar{D} + \frac{1}{6} b_t^2 \bar{D}^2 K_{iso}^{MGC} + \frac{1}{6} b_t^2 b_\Delta^2 \bar{D}^2 K_{aniso}^{MGC} + O(b^3) \tag{9}$$

Note that $K_{aniso}^{MGC}$ and $K_{iso}^{MGC}$ can be estimated from Eq. 8, using the same dMRI experiments as for CTI, since these correspond to data acquired with at least two b-tensor shapes ($b_\Delta$ = 1 for sets #1, #2, and #4, and $b_\Delta$ = -1/2 for set #3) and two non-zero b-tensor magnitudes ($b_t = b_a$ for sets #1-3, $b_t = b_b$ for set #4).

# 3. Methods

## 3.1 Simulations

The robustness of the different kurtosis source estimation strategies (MGC and CTI) was first assessed via simulations. The full details of the simulations can be found in the Supporting Information, section A. Briefly, synthetic signals were generated for different models with known ground truth kurtosis sources, which included:

1) Sum of multiple isotropic Gaussian diffusion components (Fig. 2A).

2) Sum of multiple uniformly oriented anisotropic Gaussian diffusion components with identical axial and radial diffusivities (Fig. 2B).

3) Single compartment with non-vanishing microscopic kurtosis (Fig. 2C). Although in Fig. 2C1 a single component with positive $\mu K$ is sketched as the extra-compartmental medium that encompasses randomly oriented anisotropic compartments, ground truth positive $\mu K$ in both intra- and extra-"cellular" components can arise according to effective medium theory (84) due to, e.g., cross sectional size variance and packing degree (49,85–87). The exact $\mu K$ value for a medium represented in Fig. 2C will depend on the volume fraction, anisotropy, size, and packing of the anisotropic compartments as well as on the acquisition parameters (87). To simplify, the DDE signal decay for the single isotropic compartment with positive microscopic kurtosis is here numerically computed using the signal representation $E(b_1, b_2) = \exp\left(-(b_1 + b_2)D + \frac{1}{6}(b_1^2 + b_2^2)D^2\mu K\right)$ with $D$ and $\mu K$ ground truth set to an arbitrary value of $0.65 \mu m^2/ms$ and 1, respectively. Note that $\mu K$ can also rise due to restricted diffusion. In section B of Supporting Information, simulations for diffusion inside spheres are also produced using the MISST package (88,89), which could represent neural soma (90) and other quasi-spherical objects such as boutons.

4) A system comprising different components and with non-zero contributions for all different kurtosis sources (Fig. 3A). For this system, we consider a sum of the compartment types used for the previous simulations with equal weights. As the mean diffusivities of the simulations 1, 2 and 3 are equal, this ensemble model can assess the robustness of estimates for different kurtosis sources individually by varying concrete model parameters. Given this, we then varied ground truth $K_{iso}$, ground truth $K_{aniso}$, and ground truth $\mu K$ values. Please see the Supporting Information, Section A for full details on these simulations.

For all models, powder-averaged signals were generated for the four different sets of DDE acquisition parameters (c.f. Fig. 1) for total b-values $b_a = 2.5 ms/\mu m^2$ (sets #1, #2, #3) and $b_b = 1 ms/\mu m^2$ (set #4) ($\Delta = \tau_m = 12 ms$ and $\delta = 3.5 ms$ for all experiments) – note the maximum b-value of $2.5 ms/\mu m^2$ was selected since this was showed to provide an optimal trend between signal contrast to measure diffusional kurtosis and minimization of biases from high order-terms (40). For all four sets, the 45 directions of a 3D spherical 8-design (91) were used for the single encoding of set #1 (Fig. 1, panel C1.2), for the double diffusion encodings of sets #2 and #4 (Fig. 1, panels C2.2 and C4.2), and for the first diffusion encoding of set #3 (Fig. 1, panels C3.2). The directions for the second diffusion encoding of set #3 were repeated for three equidistant orthogonal directions relative to each direction of the 3D spherical 8-design (Fig. 1, panels C3.2), yielding a total of 135 pairs of directions. To ensure homoscedasticity of powder-averaged signals, the acquisition of the 45 directions for sets #1, #2, and #4 were repeated three times (45×3=135 pairs of directions).

For reference, signals were also produced for an adapted version of the previous (*old*) DDE protocol suggested for CTI (40). The $b_1$, $b_2$ and $\theta$ values, as well as gradient directions schemes used for the improved (*new*) and *old* CTI protocols (CTI$_{new}$ and CTI$_{old}$) are summarized in Table 1 (for more information on the *old* protocol, c.f. Supporting Information, Section C). In addition to the diffusion-weighted signals for the different CTI sets, 135 signal replicas for $b_1 = b_2 = 0$ are incorporated in both protocols – this data is treated as an independent $b_t = 0$ set. To assess the robustness of estimates towards noise, all synthetic signals were corrupted by Rician noise with a nominal SNR of 40 before powder-averaging. Consideration about the precision of CTI for other SNR levels are discussed in Supporting Information, section D (c.f. Supporting Information Figure S3).

**Table 1** – Summary of the DDE parameter combination used for the "new" CTI protocol (CTI$_{new}$) and the reference "old" CTI protocol (CTI$_{old}$). Parameters b$_1$, b$_2$, b$_t$ are expressed in ms/µm².

### *New CTI protocol – CTI$_{new}$*

| set | b$_1$ | b$_2$ | b$_t$ | θ | b$_\Delta$ | direction scheme |
|---|---|---|---|---|---|---|
| #1 | 2.5 | 0 | 2.5 | 0° | 1 | 45 directions of the 8-design (x3 repetitions) |
| #2 | 1.25 | 1.25 | 2.5 | 0° | 1 | 45 directions of the 8-design (x3 repetitions) for both diffusion encodings |
| #3 | 1.25 | 1.25 | 2.5 | 90° | -1/2 | 45 directions of the 8-design for the 1$^{st}$ encoding, repeated for 3 orthogonal directions for the 2$^{nd}$ encoding |
| #4 | 0.5 | 0.5 | 1 | 0° | 1 | 45 directions of spherical 8-design (x3 repetitions) for both diffusion encodings |

### *Old CTI protocol – CTI$_{old}$*

| set | b$_1$ | b$_2$ | b$_t$ | θ | b$_\Delta$ | direction scheme |
|---|---|---|---|---|---|---|
| #1 | 2.5 | 0 | 2.5 | 0° | 1 | 45 directions of the 8-design (x3 repetitions) |
| #2 | 2.5 | 2.5 | 5 | 0° | 1 | 45 directions of the 8-design (x3 repetitions) for both diffusion encodings |
| #3 | 2.5 | 2.5 | 5 | 90° | -1/2 | 45 directions of the 8-design for the 1$^{st}$ encoding, repeated for 3 orthogonal directions for the 2$^{nd}$ encoding |
| #4 | 1 | 0 | 1 | 0° | 1 | 45 directions of the 8-design (x3 repetitions) |
| #5 | 1 | 1 | 2 | 0° | 1 | 45 directions of the 8-design (x3 repetitions) for both diffusion encodings |
| #6 | 1 | 1 | 2 | 90° | -1/2 | 45 directions of the 8-design for the 1$^{st}$ encoding, repeated for 3 orthogonal directions for the 2$^{nd}$ encoding |

<u>Data processing:</u> CTI metrics were obtained by fitting Eq. 6 to the log of the powder-averaged signals of both *new* and *old* protocols using an ordinary linear-least-squares (OLLS) procedure. $K_{aniso}$, $K_{iso}$, and $K_t$ were also estimated from the MGC approach by fitting Eq. 9 to the powder-average of CTI's *new* protocol using an OLLS procedure. Mean and standard deviation of each kurtosis estimates were computed by repeating simulations for 1000 iterations.

## 3.2. MRI experiments

All animal experiments were preapproved by the institutional and national authorities and carried out according to European Directive 2010/63. Data was acquired from N=3 female Long Evans rats (ages = 22/19/22 weeks old, weights = 354.6/260.4/334.5g in a 12 h/12 h light/dark cycle with ad libitum access to food and water) under anesthesia (~2.5% Isoflurane in 28% oxygen) using a 9.4T Bruker Biospec scanner equipped with an 86 mm quadrature transmission coil and 4-element array reception cryocoil.

Auxiliary sagittal T$_2$-weighted images were acquired using a RARE sequence (c.f. Supporting Information Section E for experimental parameters). Diffusion MRI datasets were then acquired using a DDE-EPI pulse sequence ($\Delta = \tau_m$ = 12 ms, δ = 3.5 ms) for 3 evenly spaced coronal slices placed using the sagittal T$_2$-weighted images as a reference (c.f. Supporting Information Figure S4). Per-slice respiratory gating was applied. These acquisitions followed the first CTI protocol reported in Table 1 along with 24 $b_t = 0$ acquisitions per DDE set. Other acquisition parameters included: TR/TE=3000/50.9 ms, in-plane resolution = 200×200 μm², slice thickness = 0.9 mm acquisition bandwidth = 400 kHz, number of shots = 1, partial Fourier factor = 7/8 (partial Fourier acceleration=1.25). Acquisition time for the diffusion-weighted data was around 40 mins. To empirically check if our measurements satisfied the long mixing time regime, additional DDE data was acquired in rat #1 with parallel and antiparallel directions and an intermediate total b-value of $2ms/\mu m^2$ ($b_1 = b_2 = 1ms/\mu m^2$) (Supporting Information, section F).

Data-processing: C.f. Supporting Information Section G for full details. Briefly, data were denoised (92), Gibbs unrung per channel (29,93), and the four channels were combined using sum-of-squares. Data were registered using a sub-pixel algorithm (94).

Both CTI and MGC approaches were used to provide kurtosis metrics by voxel-wise fitting of equations 6 and 9 to the set powder-averaged data (masked to avoid regions distorted due to b0 inhomogeneities (Fig. 4A) using an OLLS procedure. Moreover, for a quantitative analysis of the kurtosis sources, ten regions of interest (ROIs) were manually defined on the $b_\mathrm{t} = 0$ images of rat #1, including the left and right cortical grey matter (GM), the right and left corpus callosum genu (CCg), the right and left corpus callosum body (CCb), the right and left corpus callosum splenium (CCs), and the right and left internal capsule (IC). To ensure consistency across animals, the ROIs for rats #2 and #3 were automatically generated for the other animals using non-parametric registration of fractional anisotropy maps (95,96).

# 4. Results

## 4.1. Simulations

Figure 2 shows the simulation results for systems containing single component types (models 1-3). For isotropic Gaussian diffusion components with different mean diffusivities (Fig. 2A1), all DDE signal sets reveal identical log-signal dependencies with $b_t$ (Fig. 2A2). The non-linearity of the log-signal decays was thus correctly identified as non-zero $K_{iso}$ by all strategies (Fig. 2A3-6). For uniformly distributed anisotropic Gaussian components (Fig. 2B1), perpendicular DDE signals evidenced stronger diffusion-weighted attenuations, as expected (Fig. 2B2). The kurtosis estimated from MGC and the new CTI approaches was again correctly attributed to $K_{aniso}$ (Fig. 2B4 and Fig. 2B6). Identically biased magnitudes were observed for both strategies for $K_{aniso}$ likely due to higher-order-term biases which are expected to be larger for systems with high tissue dispersion (40). As expected, the old CTI protocol reports non-zero $\mu K$ (Fig. 2B5), while the new CTI scheme corrects this bias and attributes a zero $\mu K$ for such systems (Fig. 2B6).

For a system with microscopic disorder (Fig. 2C1, positive $\mu K$), asymmetric DDE signals (i.e., $\bar{E}_{DDE}(b_t, 0, 0°)$ ) differ from their symmetric DDE counterparts (i.e. $\bar{E}_{DDE}(b_t/2, b_t/2, 0°)$ and $\bar{E}_{DDE}(b_t/2, b_t/2, 90°)$, Fig. 2C2). The finite $\mu K$ strongly biases both $K_{aniso}$ and $K_{iso}$ from MGC (Fig. 2C4). On the other hand, CTI – both *old* and *new* protocols – correctly estimate the finite $\mu K$. These results are consistent for the non-vanishing (positive/negative) $\mu K$ values of restricted diffusion inside spheres – see Supporting Information Figure S1.

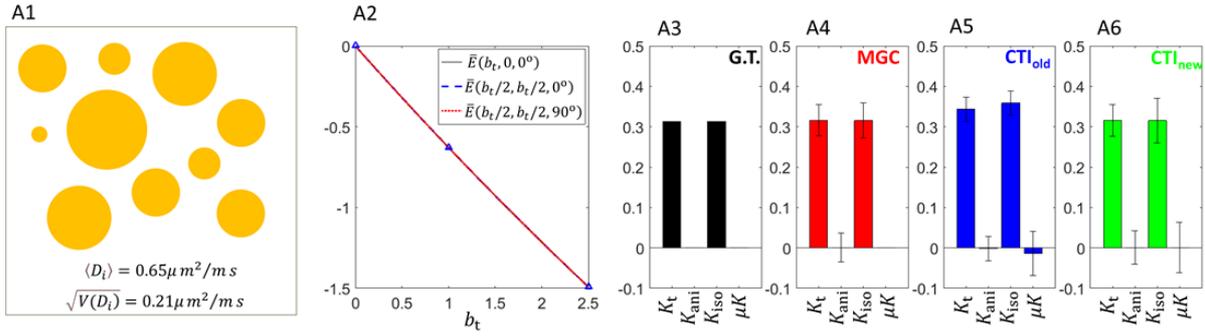

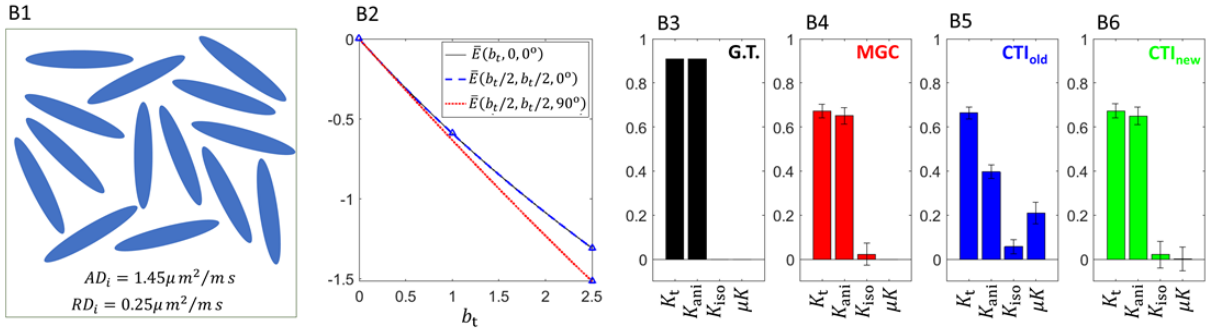

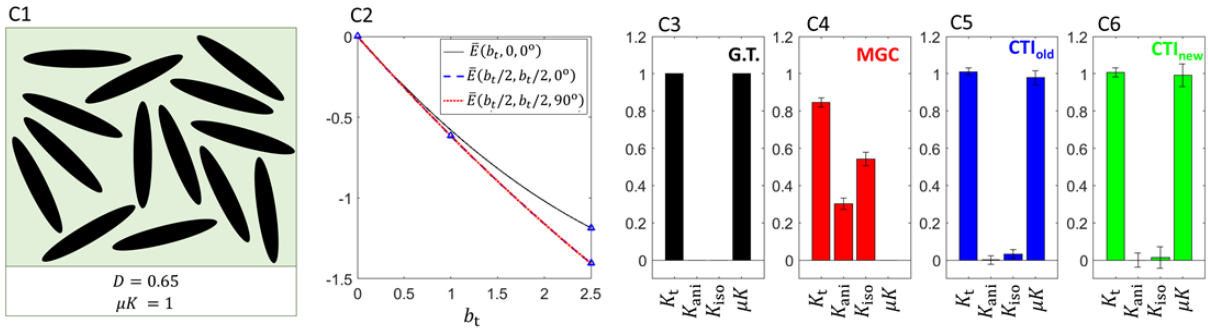

**Figure 2 – Results for synthetic diffusion-weighted signals generated according to three systems containing single compartment types: A)** Isotropic Gaussian diffusion components with different mean diffusivities sampled from a Gaussian distribution with mean $0.65 \mu m^2/ms$ and standard deviation $0.21 \mu m^2/ms$; **B)** Isotropically oriented anisotropic Gaussian diffusion components with axial and radial diffusivities of $1.45 \mu m^2/ms$ and $0.25 \mu m^2/ms$; **C)** Non-Gaussian diffusion due to microscopic disorder with ground truth $\mu K = 1$ and $D = 0.65 \mu m^2/ms$. From left to right, each panel shows: a schematic representation of the models (A1, B1, C1, D1); the signal decays for three different DDE experiment types in which signals of the improved CTI protocol are marked by the blue triangles (A2, B2, C2); the kurtosis ground truth values (A3, B3, C3); the kurtosis estimates obtained from MGC (A4, B4, C4); the kurtosis estimates obtained from the CTI using its previous "old" protocol (A5, B5, C5); the kurtosis estimates obtained from the CTI using its improved "new" protocol (A6, B6, C6). Note that when $\mu K$ exists (panel C), the MGC approaches conflate $\mu K$ with non-existing anisotropic and isotropic sources, while CTI clearly identifies the $\mu K$ in the system. Also note that the new CTI protocol better resolves anisotropic and isotropic sources compared to its old counterpart (panels A and B).

In a realistic voxel, all kurtosis sources could exist simultaneously, and it is therefore instructive to assess whether the CTI framework can disentangle the kurtosis sources with specificity when all sources are present (Fig. 3). For this system, log-signal DDE decays are different for the three conditions (Fig. 3A2). The *new* CTI protocol successfully estimates the kurtosis sources (Fig. 3A3-6), while the *old* protocol overestimates $\mu K$.

We then investigated how changes in ground truth kurtosis sources would impact the different source estimates (see methods). Particularly, when changing the ground truth $K_{iso}$ (Fig. 3B) or $K_{aniso}$ (Fig. 3C), the new CTI and MGC approaches correctly show larger changes in $K_{iso}$ and $K_{aniso}$ respectively. The *old* CTI protocol has limited specificity for µK as varying $K_{aniso}$ clearly affects $\mu K$. $K_{aniso}$ and $K_{iso}$ for the MGC approach also show limited specificity when the ground truth $\mu K$ is varied (Fig. 3D).

Interestingly, we find that CTI *new* protocol correctly tracked the specific ground truth sources, i.e. major changes in improved CTI $K_{iso}$, $K_{aniso}$, and $\mu K$ estimates are only observed when $K_{iso}$, $K_{aniso}$, and $\mu K$ ground truths are varied, respectively (Fig. 3B3, Fig. 3C2, and Fig. 3D4), even if some offsets exist. Note that in Figure 3, due to the biases in MGC estimates introduced by $\mu K$, these only match the improved CTI estimates when ground truth $\mu K$ is zero (c.f. Fig. 3D).

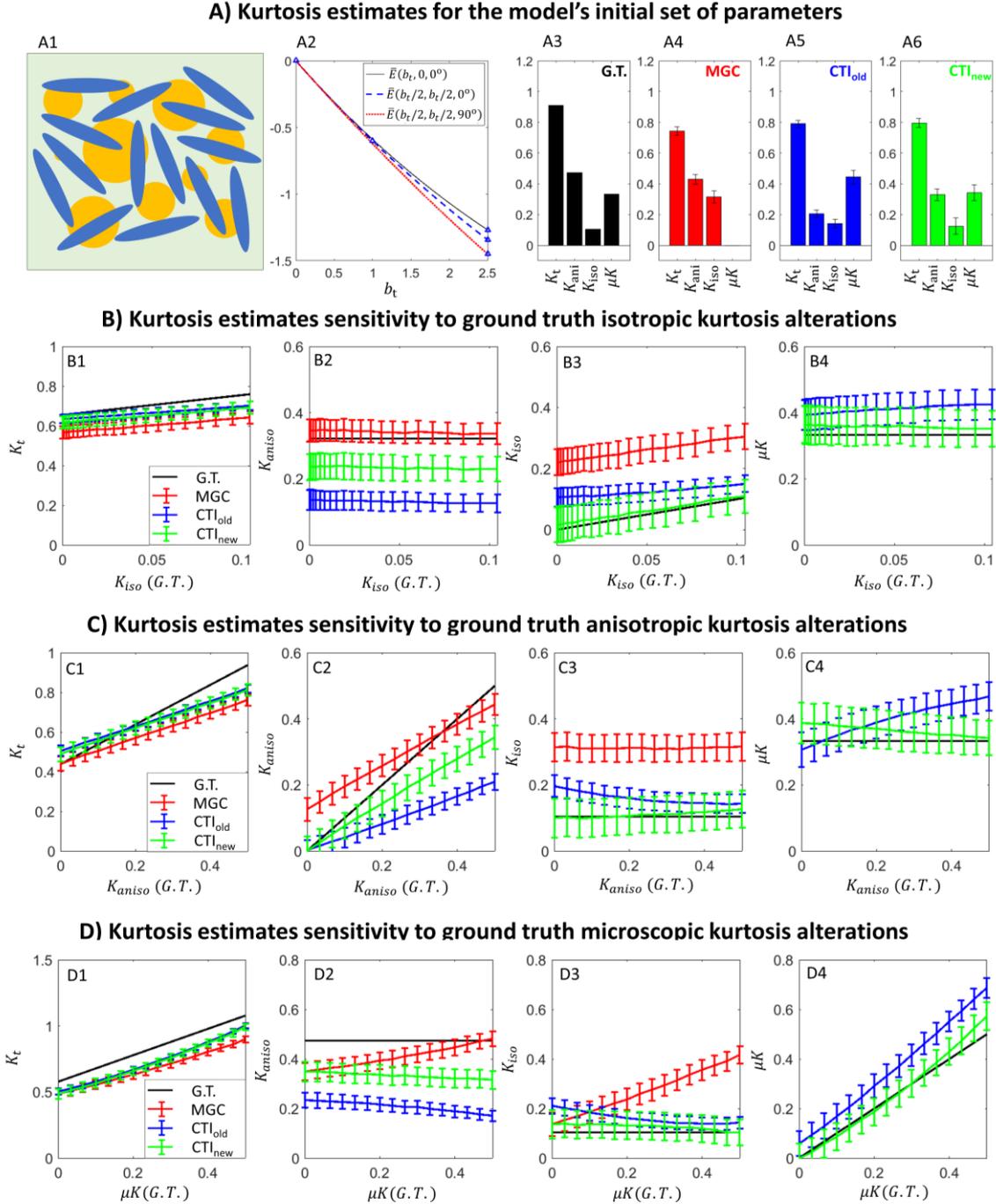

**Figure 3 – Results for synthetic diffusion-weighted signals generated according to a system containing a sum of different compartment types (isotropic Gaussian + anisotropic Gaussian + isotropic non-Gaussian components):** **A)** Results from the model's initial guess, i.e., isotropic components' mean diffusivities sampled from a Gaussian distribution with mean $0.65 \mu m^2/ms$ and standard deviation $0.21 \mu m^2/ms$, anisotropic Gaussian diffusion components with axial and radial diffusivities of $1.45 \mu m^2/ms$ and $0.25 \mu m^2/ms$; and non-Gaussian diffusion with $K_i = 1$ and $D_i = 0.65 \mu m^2/ms$. **B)** CTI and MGC kurtosis estimates' sensitivity to ground truth $K_{iso}$ alterations by changing the mean diffusivity variation of isotropic Gaussian components. **C)** CTI and MGC kurtosis estimates sensitivity to ground truth $K_{aniso}$ alterations by changing the axial and radial diffusivities of anisotropic Gaussian compartments. **D)** CTI and MGC estimates sensitivity to ground truth $\mu K$ alterations by changing the microscopic kurtosis level of the non-Gaussian component. These simulations suggest a very good stability for CTI (especially the new protocol), such that a change in one direction of any of the metrics will induce a systematically correct change in the estimated values.

## 4.2. MRI experiments

Coronal $b_t = 0$ images for all three rats are shown in Fig. 4A. Nominal SNR was ~30 for the raw data (Fig. 4B). Fig. 4C shows a representative diffusion-weighted image for a dMRI experiment acquired for the maximum b-value used on this study before and after PCA denoising. An SNR gain of ~1.3 was noted upon denoising (the nominal SNR of the denoised data was ~40).

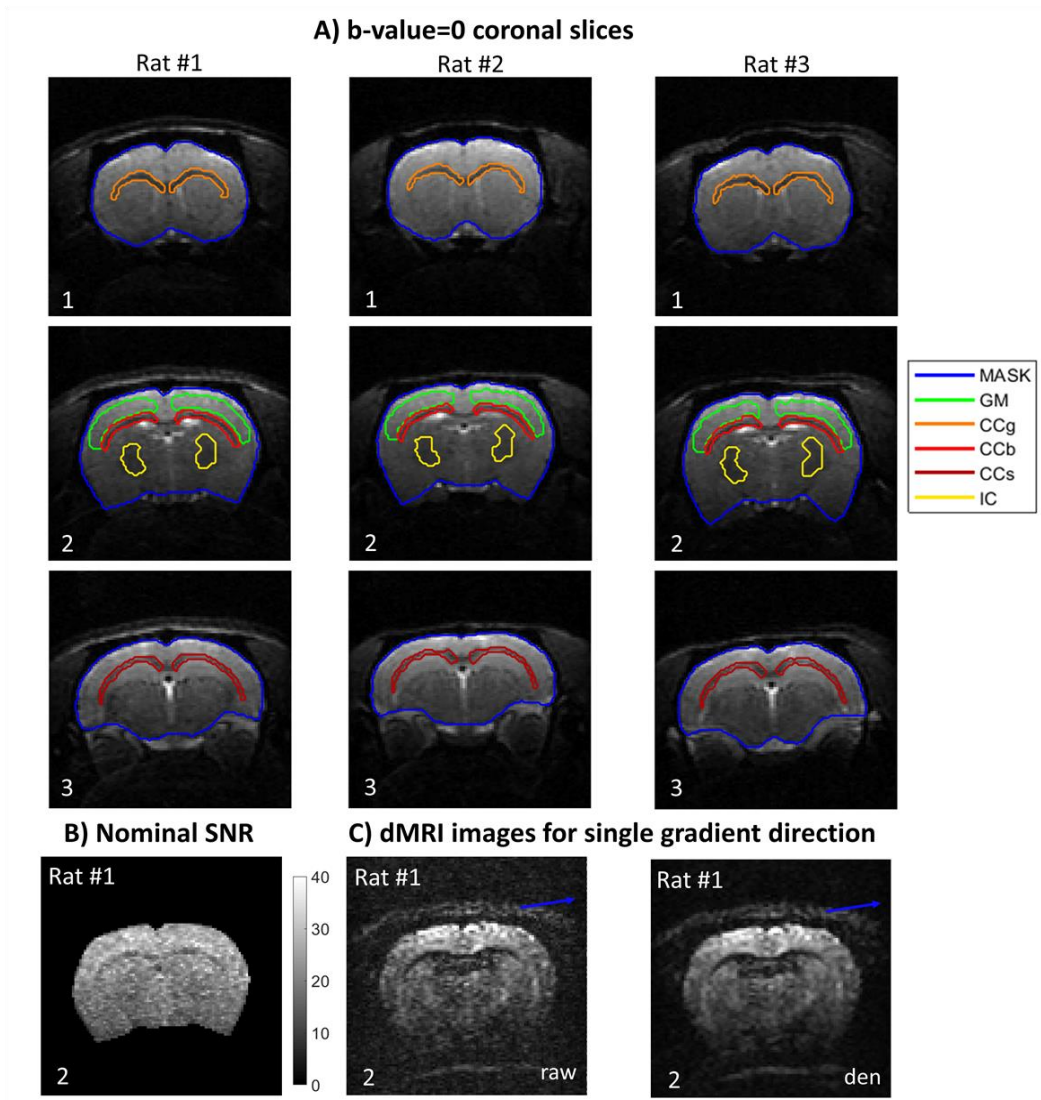

**Figure 4 – Raw diffusion-weighted data: A)** Representative $b_t = 0$ image for all animals and coronal slices and delineate regions of interest: brain mask (MASK); right and left cortical grey matter (GM); right and left corpus callosum genu (CCg); the right and left corpus callosum body (CCb); right and left corpus callosum splenium (CCs); and right and left internal capsule (IC). **B)** Representative nominal SNR map that was computed from all $b_t = 0$ acquisitions of Rat #1 and slice 2. **C)** Representative diffusion-weighted images (Rat #1, slice 2) for a single diffusion gradient direction (marked in blue) for the maximum total $b_1$ used in this study (i.e. $b_1 = 2.5\ \mu m^2/ms$ and $b_2 = 0$) before and after PCA denoising (right and left respectively). Note the high signal to noise, and the lack of imaging artifacts.

Fig. 5A shows the powder-averaged signal decays for the four sets of the improved CTI protocol. The log difference between the powder-averaged data from set #1 and set #2 (Fig. 5B1) shows the sensitivity to $\mu K$ (c.f. Eq. 4), while the log difference between the powder-averaged data from set #2 and set #3 (Fig. 5B4) shows CTI's sensitivity to $K_{aniso}$ (c.f. Eq. 5). Positive log difference between set #1 and set #2, reveals a non-vanishing positive $\mu K$ for both grey and white matter (c.f. Fig. 5B2 and Fig. 5B3).

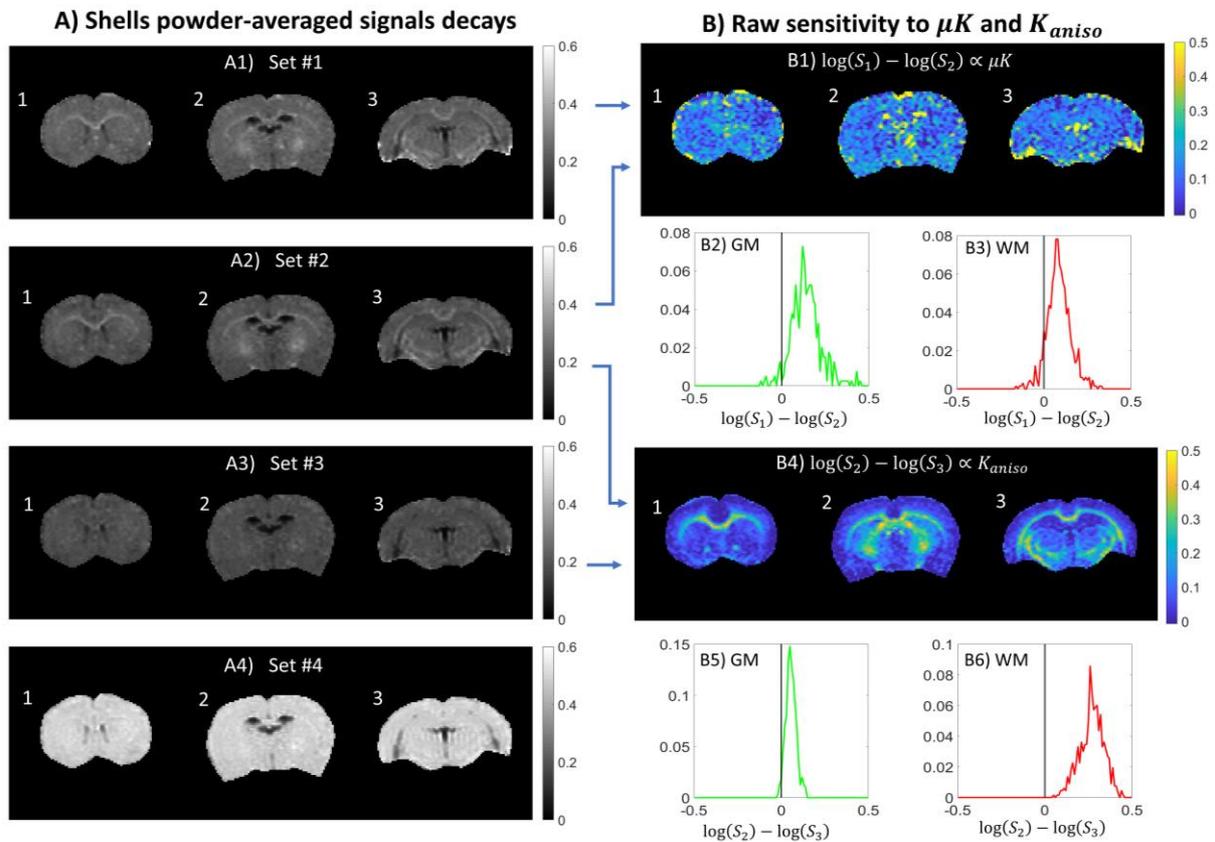

**Figure 5 – Results from powder-averaged data for all slices of a representative animal (Rat #1). A)** Powder-averaged signal decays for all four sets DDE experiments for CTI: set #1 (A1); set #2 (A2); set #3 (A3); and set #4 (A4); **B)** Raw sensitivity to microscopic kurtosis ($\mu K$) and anisotropic kurtosis ($K_{aniso}$): raw $\mu K$ sensitivity maps are quantified by the log difference between powder-averaged signals of set #1 and #2 (B1); histogram of the log difference between powder-averaged signals of set #1 and #2 for all concatenated grey matter (GM) and white matter (WM) regions of interest (B2-3); raw $K_{aniso}$ sensitivity maps are quantified by the log difference between powder-averaged signals of set #2 and #3 (B4); histogram of the log difference between powder-averaged signals of set #2 and #3 for all concatenated grey matter (GM) and white matter (WM) regions of interest (B5-6). Note that the data have sufficient power to resolve $\mu K$ and $K_{aniso}$ even for raw data itself.

The CTI kurtosis estimates for all slices of rat #1 and the second slice of rats #2 and #3 are shown in Fig. 6. Both $K_t$ and $K_{aniso}$ were higher in white matter regions (Fig. 6A-B). $K_{iso}$ and $\mu K$ maps show noisier spatial profiles than $K_t$ and $K_{aniso}$ maps (Fig. 6C-D). Nevertheless, $\mu K$ shows to be a prevalent source of kurtosis, presenting higher values in grey matter and lower values in white matter (Fig. 6D, red arrows). The ROI analysis supported the trends observed in the kurtosis maps (Fig. 7A) – in general, $\mu K$ shows to explain 64±6% and 30±14% of the total kurtosis in grey and white matter regions, respectively (Fig. 7, panel A4). The higher standard deviations for $K_{iso}$ and $\mu K$ is consistent with the lower precision observed in their maps (standard deviation of $\mu K$ estimates are in line to the estimation error predicted in Supporting Information Figure S3). Fig. 7B shows the histograms of the different CTI-driven kurtosis estimates for combined grey and combined white matter ROIs. The mean values of grey and white matter are significantly different (two-sample t-test with unequal variances, p<0.001 for kurtosis estimates and all three animals).

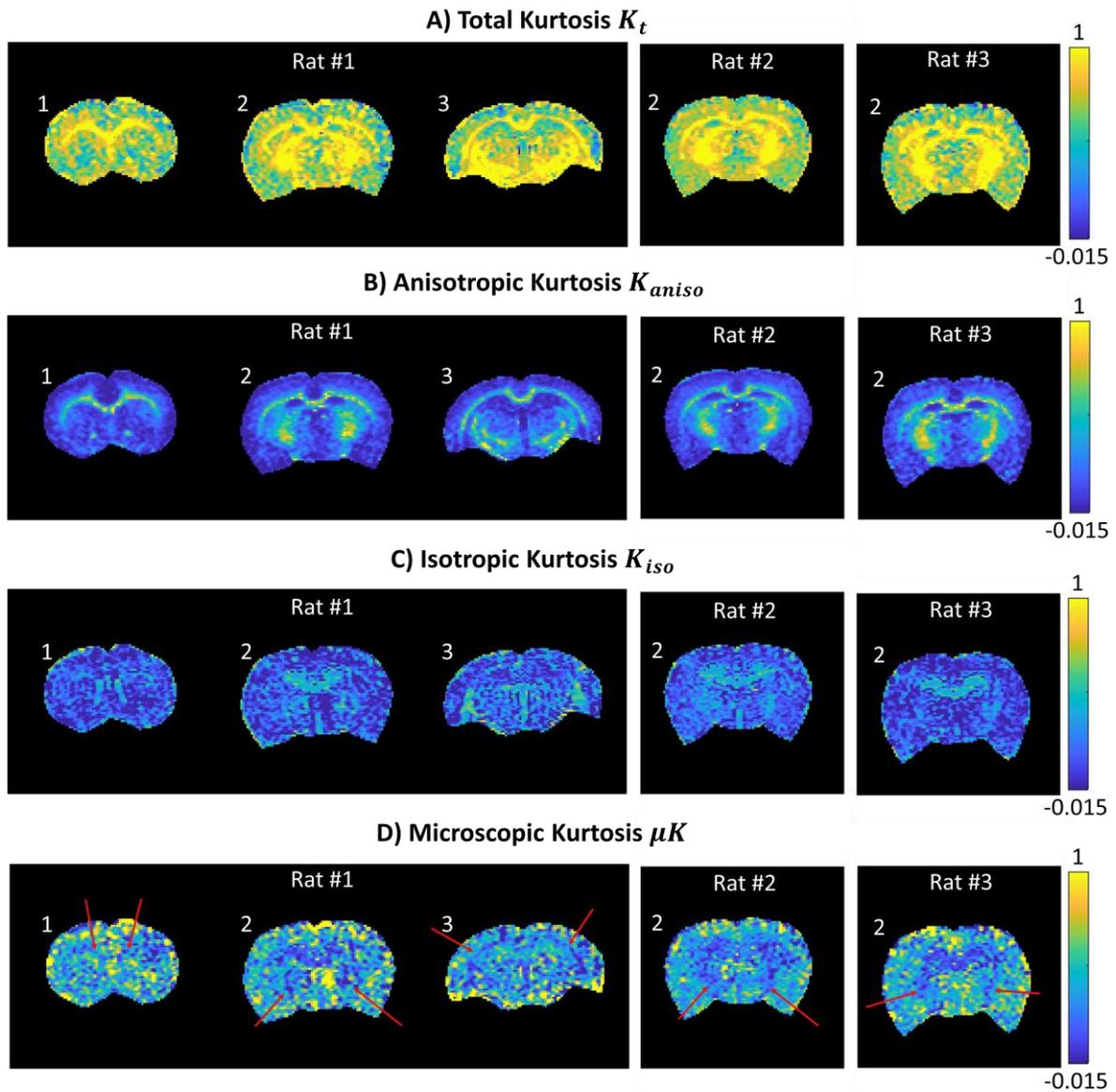

**Figure 6 – Correlation Tensor Imaging kurtosis estimates. A)** Total Kurtosis $K_t$; **B)** Anisotropic Kurtosis $K_{aniso}$; **C)** Isotropic Kurtosis $K_{iso}$; **D)** Microscopic Kurtosis $\mu K$. The red arrow in panel D points to lower $\mu K$ estimates observed particularly in white matter brain regions. On each panel, maps are presented for all slices of rat #1 and the 2nd slice of rats #2 and #3. Note how the total kurtosis maps, which have the largest values, are decomposed into their underlying sources. Anisotropic kurtosis is highest in WM, while isotropic kurtosis is highest in areas with larger partial volume effects between different diffusivities (namely areas with cerebrospinal fluid). Although the µK maps are slightly noisier, they do display GM/WM contrast.

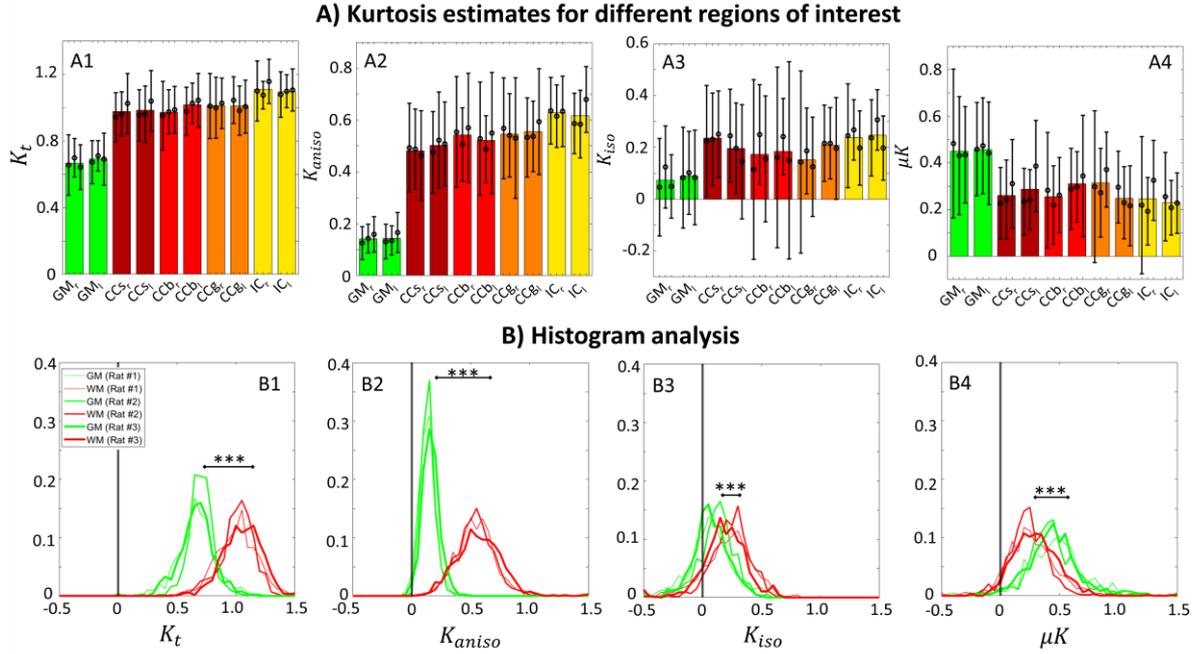

**Figure 7 – Correlation Tensor Imaging kurtosis estimates in various brain regions. A)** Mean and standard deviation of $K_t$ (A1), $K_{aniso}$ (A2), $K_{iso}$ (A3), and $\mu K$ (A4) estimates for different regions of interest – the three black intervals on each bar shows the mean and standard deviations for the three individual rats, while the coloured bars show the mean values across all three animals. Regions of interest for these panels include: the right and left cortical grey matter ($GM_r$ and $GM_l$); the right and left corpus callosum splenium ($CCs_r$ and $CCs_l$); the right and left corpus callosum body ($CCd_r$ and $CCd_l$); the right and left corpus callosum genu ($CCg_r$ and $CCg_l$); and the right and left internal capsule ($IC_r$ and $IC_l$). **B)** Histograms of the $K_t$ (B1), $K_{aniso}$ (B2), $K_{iso}$ (B3), and $\mu K$ (B4) estimates for grey matter (histograms in green) and white matter (histograms in red) regions of interest - in each panel, the differences between the mean values from grey and white matter ROIs are statistically tested using a two-sampled t-test with unequal variances (* for p<0.05, ** for p < 0.01, and *** for p<0.001 ***). The histograms reveal that µK is in fact the dominant contributor to the total kurtosis in the GM.

We then turned to assess the impact of the finite $\mu K$ on multiple gaussian component (MGC) analysis using only diffusion tensor variance. Figure 8A-C shows the kurtosis maps obtained from the MGC approach, while Figure 8D shows the histograms of MGC kurtosis values from white and grey matter ROIs. In comparison to their CTI counterparts, MGC-derived $K_t$ was lower in both grey and white matter regions (Fig. 8A), while MGC $K_{aniso}$ and $K_{iso}$ values were higher. As for CTI, MGC mean $K_t$ and $K_{aniso}$ appear higher in white matter; however, non-significant differences between the white and grey matter voxels were observed for the MGC $K_{iso}$ estimates (Fig. 8D3).

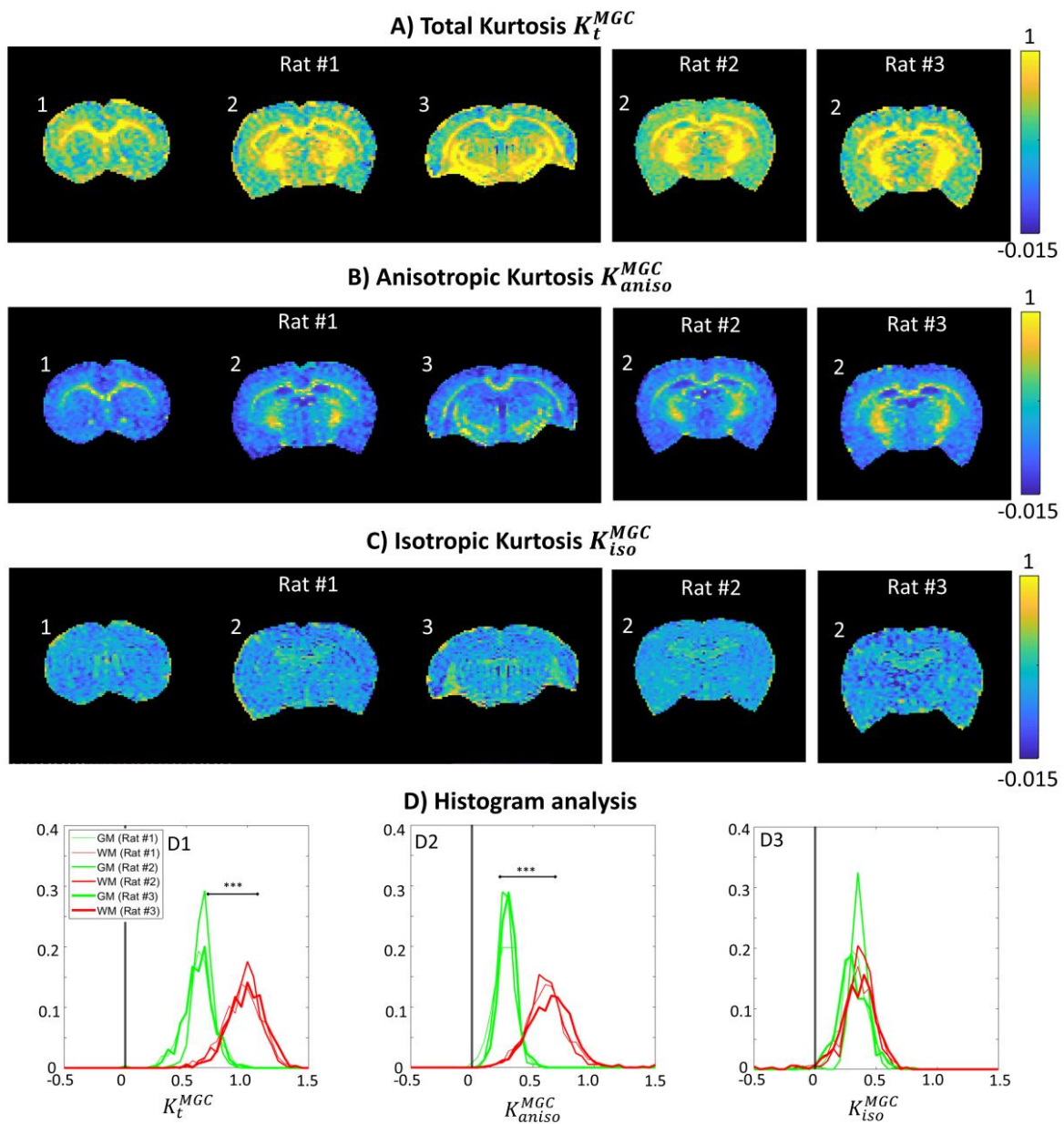

**Figure 8 – Kurtosis estimates assuming multiple gaussian components (MGC). A)** MGC total kurtosis ($K_t^{MGC}$) maps. **B)** MGC anisotropic kurtosis $K_{aniso}^{MGC}$ maps. **C)** MGC isotropic kurtosis $K_{iso}^{MGC}$ maps. **D)** Histograms of the $K_t$ (D1), $K_{aniso}$ (D2), and $K_{iso}$ (D3) MGC estimates for all grey matter (histograms in green) and white matter (histograms in red) regions of interest – in each panel, the differences between the mean values of grey and white matter ROIs are statistically tested using a two-sampled t-test with unequal variances (* for p<0.05, ** for p < 0.01, and *** for p<0.001 ***). When forcing an MGC analysis on the data, the isotropic and anisotropic kurtosis metrics "absorb" the ignored µK, leading to biased maps (compare the MGC-driven maps in this Figure with those shown in Figure 6A-C, respectively, and the histograms in this figure with those shown in Figure 7B).

To further investigate the correlation between these metrics, scatter plots of MGC and CTI kurtosis estimates are shown in Fig. 9. Points in the scatter plots are color-coded according to CTI's microscopic kurtosis ($\mu K$) estimates, showing that higher differences between CTI and MGC estimates are associated with higher degrees of $\mu K$.

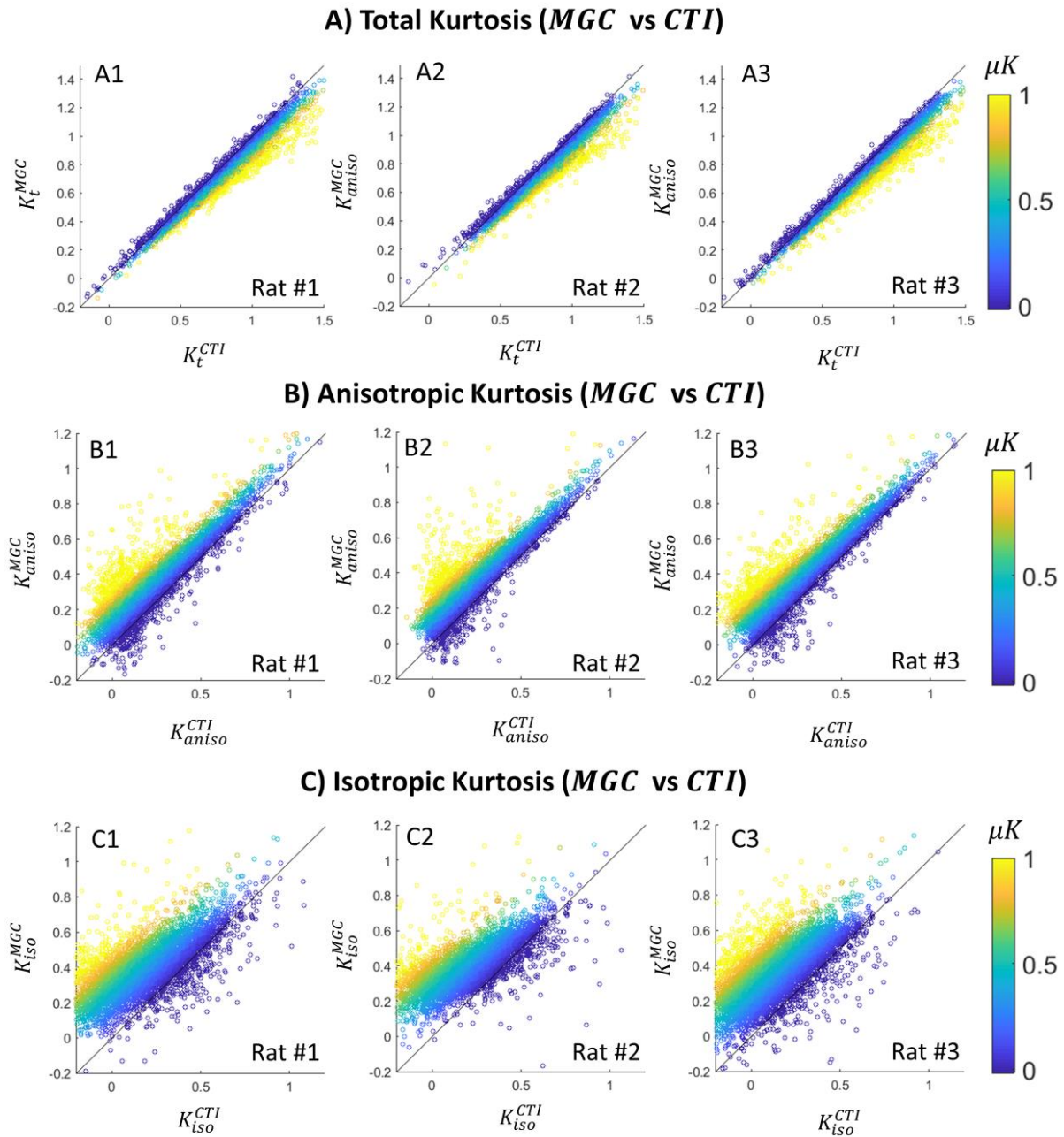

**Figure 9 – Scatter plots between MGC and CTI estimates. A)** Scatter plots between MGC total kurtosis ($K_t^{MGC}$) and CTI total kurtosis ($K_t^{CTI}$) estimates; **B)** Scatter plots between MGC anisotropic kurtosis ($K_{aniso}^{MGC}$) and CTI anisotropic kurtosis ($K_{aniso}^{CTI}$) estimates; **C)** Scatter plots between MGC isotropic kurtosis ($K_{iso}^{MGC}$) and CTI isotropic kurtosis ($K_{iso}^{CTI}$) estimates. Points in the scatter plots are colour coded according to CTI's microscopic kurtosis ($\mu K$) estimates. Note that as μK increases, the bias in the MGC estimates becomes more dramatic, especially for isotropic and anisotropic kurtosis sources.

# 5. Discussion

The conflation of underlying kurtosis sources in SDE was a major motivation in developing multidimensional diffusion encoding approaches. Even under the multiple gaussian component assumption, $K_{aniso}$ and $K_{iso}$ contrasts have shown great promise for e.g. distinguishing different tumor types and grades (39,72), depicting healthy and pathological age-related microstructural alterations (97), mapping multiple sclerosis lesions (98,99), and characterizing body organs (69,100). However, the MGC assumption implicitly ignores diffusion-time dependence (73,74) and $\mu K$ effects, thereby risking the conflation of $\mu K$ effects into the metrics. We therefore sought in this study to characterize the insofar ignored $\mu K$ *in vivo* and assess its impact on the more conventional MGC approaches.

The Correlation Tensor Imaging (CTI) approach was recently introduced for $\mu K$ mapping. CTI goes beyond the tensor-valued framework and simultaneously estimates $K_{aniso}$, $K_{iso}$, and $\mu K$ – albeit at the expense of a larger number of acquisitions (40). The original CTI framework was lengthy and included DDE measurements that could bias $\mu K$ in some scenarios (e.g., Fig. 2B5) due to higher order terms – a common issue for techniques based on the truncation of the signal cumulant expansion (40,77,101). To alleviate these drawbacks and accelerate the acquisitions, a new CTI strategy was here developed. We found a sparser set of DDE acquisitions for robustly resolving kurtosis sources and found that they can much more accurately estimate $\mu K$ compared with the older protocol. By balancing the total b-values used for different DDE sets (c.f. Section C of the Supporting Information), the higher order term effects are greatly diminished (e.g. Fig. 2B6 and Fig. 3). The four CTI quantities ($\bar{D}$, $K_{aniso}, K_{iso}, \mu K$) can be fully resolved from only four different combinations of DDE parameters ($b_1, b_2, \theta$) (c.f. DDE the sets in Fig. 1). In addition, we managed to accelerate CTI from 2h to under 40 mins, thereby making it applicable for *in vivo* preclinical and even clinical mapping. It is interesting to note that $\mu K$ alone can be estimated from the log signal differences of two different DDE experiments (c.f. Eq. 7), specifically using 1) parallel symmetric DDE gradient waveforms and 2) experiments analogous to SDE with the same total b-value, much like how microscopic anisotropy is estimated from parallel and perpendicular waveforms (54,55,57,76,77)). Therefore, experiments aiming to resolve only $\mu K$ could be even further accelerated.

In (40), we found significant positive $\mu K$ estimates both in grey and white matter brain regions in the *in vivo* rat brain, suggesting it cannot be ignored in MGC approaches. Our new results with the improved CTI protocol confirmed the overall non-vanishing positive $\mu K$ effects in both white and grey matter (Fig. 6, Fig. 7) and further highlighted significant $\mu K$ differences between grey and white matter brain regions (Fig. 7). Positive $\mu K$ is consistent with non-Gaussian diffusion effects due to intra-cellular cross-section size variance (18,49,86) and/or the presence of obstacles in tortuous extra-cellular environments (84,87). Therefore, the $\mu K$ differences between grey and white matter could perhaps be explained by differences in both intra- and extra-cellular microstructural configurations (e.g. different compartmental cross sectional variance, different degree of cellular packing, etc.), or by the presence of a more negative $\mu K$ contributions from restricted diffusion in white matter. Although $\mu K$ will depend on a weighted sum of all above-mentioned effects, positive $\mu K$ contributions are expected to prevail over negative $\mu K$ contributions from completely restricted diffusion, as the latter are typically associated with low apparent diffusivities which strongly affects the signal contributed from these compartments (note that the total $\mu K$ measured by CTI is a weighted average of all its contributions, where the weights depend on the squared apparent diffusivities of each contribution ($\mu K = \frac{\langle D_i^2 \mu K_i \rangle}{\overline{D}^2}$)).

Another significant result in this study, is that this non-vanishing $\mu K$ can have a dramatic effect on kurtosis sources computed from tensor-valued and MGC framework (Fig. 8 and Fig. 9). In general, both $K_{aniso}$ and $K_{iso}$ derived from the MGC approach were biased towards higher values compared with their more accurate CTI counterparts. The color-coded scatter plots between MGC and CTI estimates revealed that differences can be fully explained by $\mu K$ biases on MGC estimates (Fig. 9). In is important to note, that the influence of $\mu K$ on $K_{aniso}$ and $K_{iso}$ can obscure microstructural differences, such as the $K_{iso}$ differences between grey and white matter brain regions (Fig. 9).

## 5.1. Limitations and future work

Although it does not rely on the Gaussian diffusion assumption, CTI is still a cumulant expansion of DDE signals, which induces some implicit assumptions. Namely, disregarding higher-order cumulant terms; assuming the long mixing time regime (which can be empirically evaluated); and ignores exchange. Higher-order-term biases were here minimized by the new CTI protocol. The long mixing time regime effectively suppresses unwanted time-dependent diffusion correlations from the Q and S-tensors (60,80) and guarantees that the Z-tensor in Eq. 5 approaches the covariance tensor (40,60,76,80). This regime was empirically verified in acquired signals by measuring parallel and antiparallel DDE experiments and showing that they produce identical signal decays (55,57,59,102,103) (c.f., Supporting Information Figure S5). Lastly, since exchange is not explicitly modelled by CTI, it may affect the microscopic kurtosis metrics (19,42,104,105). Future studies will investigate the biological underpinning of microscopic kurtosis and how exchange between biological components can affect its estimates.

In this study, simple models were used to investigate the origin of CTI-driven kurtosis sources, and to illustrate the impact of finite $\mu K$ on previous MGC approaches. These simple models are, however, are likely not sufficiently complex to fully represent biological tissues. Future studies should expand such in-silico experiments toward more complex simulations allowing the assessment of the relationships between different kurtosis sources and concrete (sub)cellular features (e.g., cellular cross-sectional variance, cellular packing, exchange, etc.) (86,104,106–109). Another limitation of the current study is the reduced number of animals used. Although the power of measuring non-vanishing $\mu K$ is ensured by using large ROIs (i.e., statistical power ~100% considering that $\mu K$ is around 0.26 and 0.45 for ROIs containing more than 80 voxels and assuming that measures have a precision of ~0.2), future studies using a larger number of animals could be performed to explore for detailed regional differences of microscopic kurtosis.

Lundell et al. (2019) have recently shown that continuous diffusion gradient waveforms experiments probing identical b-tensors but different power spectra can provide different information about diffusion time-dependence and microscopic kurtosis (67). One could therefore argue MGC-driven biases could be reduced by adjusting the diffusion parameters of our CTI acquisition protocol, or by entirely excluding some acquisitions. In section H of Supporting Information, we report MGC kurtosis estimates obtained by fitting Eq. 9 to only DDE sets #2, #3, and #4, which correspond to an acquisition with identical waveform profiles. Under this condition, MGC $K_{aniso}$ and CTI $K_{aniso}$ are identical (c.f., Supporting Information Figure S6). However, the MGC $K_{iso}$ estimates from this modified protocol are still a combination of isotropic and microscopic kurtosis effects. The comparison between CTI and MGC approaches could assist to define the regimes in which multidimensional MGC estimates are accurate.

Here, we managed to accelerate the CTI scan times to about 40 mins. Although this is still not (yet) sufficiently rapid for clinical translation, we note that the objective was only to identify the minimal acquisition set requirements for the extraction of all CTI quantities. Indeed, we acquired a large number of directions (135 per experiment set) to enhance the precision of our kurtosis estimates. In future studies, further acceleration could be obtained by reducing the number of directions acquired for powder-averaging (78,99). We note that microscopic kurtosis can also be estimated directly from only 2 sets of this minimal protocol – albeit at the expense of not resolving $K_{aniso}$ and $K_{iso}$ – as was done for the raw $\mu K$ sensitivity analysis in Fig. 5B. Future studies aiming to measure $\mu K$ should also consider the desired estimation precision/accuracy when designing their experiments – some considerations on the relationship between $\mu K$ precision and acquisition parameters are described in section D of the Supporting Information.

# 6. Conclusion

The accelerated Correlation Tensor Imaging approach developed here facilitated more accurate in-vivo acquisitions, and revealed that the commonly neglected $\mu K$ is a significant source of total kurtosis in the brain, both in grey matter and in white matter. In fact, $\mu K$ is the dominant kurtosis source in grey matter. Ignoring $\mu K$ leads to significant bias in MGC approaches, underscoring the importance of accounting for $\mu K$ in multidimensional diffusion encoding approaches. Our findings suggest promising new biomarkers in health and disease.

# Acknowledgments

This study was funded by the European Research Council (ERC) (agreement No. 679058). The authors acknowledge the vivarium of the Champalimaud Centre for the Unknown, a facility of CONGENTO which is a research infrastructure co-financed by Lisboa Regional Operational Programme (Lisboa 2020), under the PORTUGAL 2020 Partnership Agreement through the European Regional Development Fund (ERDF) and Fundação para a Ciência e Tecnologia (Portugal), project LISBOA-01-0145-FEDER-022170.

# Supporting Information

**Section A – Simulation details (Methods)**

To assess the robustness of the different kurtosis source estimation strategies (MGC and CTI), we harnessed details simulations for the following scenarios:

5) Sum of multiple isotropic Gaussian diffusion components with mean diffusivities $D_i$ sampled from a Gaussian distribution with mean $\langle D_i \rangle = 0.65 \mu m^2/ms$ and standard deviation $\sqrt{V(D_i)} = 0.21 \mu m^2/ms$ (Fig. 2A). Ground truth parameters for this model are: $D = 0.65 \mu m^2/ms$, $K_t = K_{iso} = 0.31$, $K_{aniso} = \mu K = 0$.

6) Sum of multiple uniformly oriented anisotropic Gaussian diffusion components all with the same axial and radial diffusivities of $AD_i = 1.45 \mu m^2/ms$ and $RD_i = 0.25 \mu m^2/ms$ (Fig. 2B). Ground truth parameters for this model are: $D = 0.65 \mu m^2/ms$, $K_t = K_{aniso} = 0.91$, $K_{iso} = \mu K = 0$.

7) Single compartment with positive source of microscopic kurtosis. According to the effective medium theory (84), compartments with microscopic disorder can exhibit non-Gaussian diffusion with a positive microscopic kurtosis contribution $\mu K_i$ (84,85)). Although in Fig. 2C a single component with positive $\mu K$ is sketched as an extra-compartmental medium that encompasses randomly oriented anisotropic compartments, microstructural disorder in both intra- and extra-"cellular" components arising, e.g., due to cross sectional size variance and packing degree (49,85–87). The exact $\mu K$ value for a medium represented in Fig. 2C will depend on the volume fraction, anisotropy, size, and packing of the anisotropic compartments as well as on the acquisition parameters (87). To simplify, the DDE signal decay for the single isotropic compartment with positive microscopic kurtosis is here numerically computed using the signal representation $E(b_1, b_2) = \exp\left(-(b_1 + b_2)D + \frac{1}{6}(b_1^2 + b_2^2)D^2 \mu K\right)$ with $D$ and $\mu K$ ground truth set to an arbitrary value of $0.65 \mu m^2/ms$ and 1, respectively.

8) A system comprising different components and with non-zero contributions for all different kurtosis sources (Fig. 3A). For this system, we consider a sum of the compartment types used for the previous simulations with equal weights. The signal for this model is first computed with the diffusivity values specified above, resulting in the following ground truth parameters: $D = 0.65 \mu m^2/ms$, $K_t = 0.911$, $K_{aniso} = 0.473$, $K_{iso} = 0.104$, and $\mu K = 0.333$. As the mean diffusivities of the simulations 1, 2 and 3 are equal, this ensemble model can assess the robustness of estimates for different kurtosis sources individually by varying concrete model parameters. In particular, different ground truth $K_{iso}$ were generated by changing the mean diffusivity variance of Gaussian isotropic components $K_{iso}^{(g.t.)} = 3f_1 V(D_i^{(1)})/D^2$; different ground truth $K_{aniso}$ were created by changing the difference between the axial and radial diffusivities ($\alpha = AD_i^{(2)} - RD_i^{(2)}$) of uniformly oriented anisotropic Gaussian components $K_{aniso}^{(g.t.)} = 4f_2 \alpha^2/15D^2$. In the latter case, $AD_i^{(2)}$ and $RD_i^{(2)}$ were computed as $D + 2\alpha/3$ and $D - \alpha/3$ for $\alpha$ values sampled from 0 to $1.95 \mu m^2/ms$ to keep the mean diffusivity constant. Finally, different ground truth $\mu K$ values were generated by changing the microscopic kurtosis contribution of the non-Gaussian diffusion compartment $\mu K^{(g.t.)} = (1 - f_1 - f_2)\mu K^{(3)}$.

## Section B – CTI in spherical compartments

The signals for restricted diffusion are produced using the MISST package (88, 89). These signals are first produced for a compartment radius of 3 μm and intrinsic diffusivity of 2 μm²/ms (Supporting Information Figure S1 panel A). Note that these first simulations may represent, for example, small neural soma e.g. (90), and other smaller abundant quasi-spherical objects such as boutons.

In analogous to the system with microscopic disorder (Fig. 2C1), asymmetric DDE signals (i.e., $\bar{E}_{DDE}(b_t, 0, 0°)$ ) for restricted diffusion differ from their symmetric DDE counterparts (i.e. $\bar{E}_{DDE}(b_t/2, b_t/2, 0°)$ and $\bar{E}_{DDE}(b_t/2, b_t/2, 90°)$, Supporting Information Fig. S1A2). Since these simulations comprise of a single isotropic compartment (where $K_t = \mu K$, and the other sources are identically zero), the respective microscopic kurtosis ground truth can be calculated from single diffusion encoding at single direction and multiple b-values using DKI. For this, synthetic signals for 26 evenly sampled b-values between 0 and $2.5 ms/\mu m^2$ are generated (for $\Delta = 12ms$ and $\delta = 3.5ms$). Ground truth $K_t$ and $\mu K$ are shown in Supporting Information Fig. S1A3. MGC and CTI kurtosis estimates from the improved and old protocols are shown in Supporting Information Fig. S1A4-6. These panels show that CTI correctly estimate the finite $\mu K$, while the presence of negative $\mu K$ biases both $K_{aniso}$ and $K_{iso}$ from MGC.

Restricted diffusion simulations are then expanded to spherical compartment with varying angles. Panels B.1 and B.3 of Supporting Information Figure S1 shows the ground truths mean diffusivities and microscopic kurtosis as a function of the radius of the compartment. Synthetic signals for the CTI fitting are then sampled according to the improved protocol described on the main manuscript ($b_a = 2.5 ms/\mu m^2$, $b_b = 1 ms/\mu m^2$, $\Delta = t_m = 12ms$ and $\delta = 3.5ms$). To check for the long mixing time regime, we compare these signals with the signals generated by inverting the second gradient direction of the DDE experiments. Signal discrepancy between parallel and antiparallel signals were observed for spherical compartments for $r > 3\mu m$. Thus, to suppress diffusion time dependent effects on DDE signals, geometrically averaged signals are computed as $S_g = \sqrt{S_p S_a}$, where $S_p$ and $S_a$ are the signals of the improved protocol with non-inverted and inverted second gradient direction (note that this correction is not performed on the results of the main manuscript since the long mixing time regime was empirically shown to hold – see supporting information, section F). CTI mean diffusivities and microscopic kurtosis computed from noise free geometrically averaged signals are shown in panels B2 and B4 of Supporting Information Figure S1.

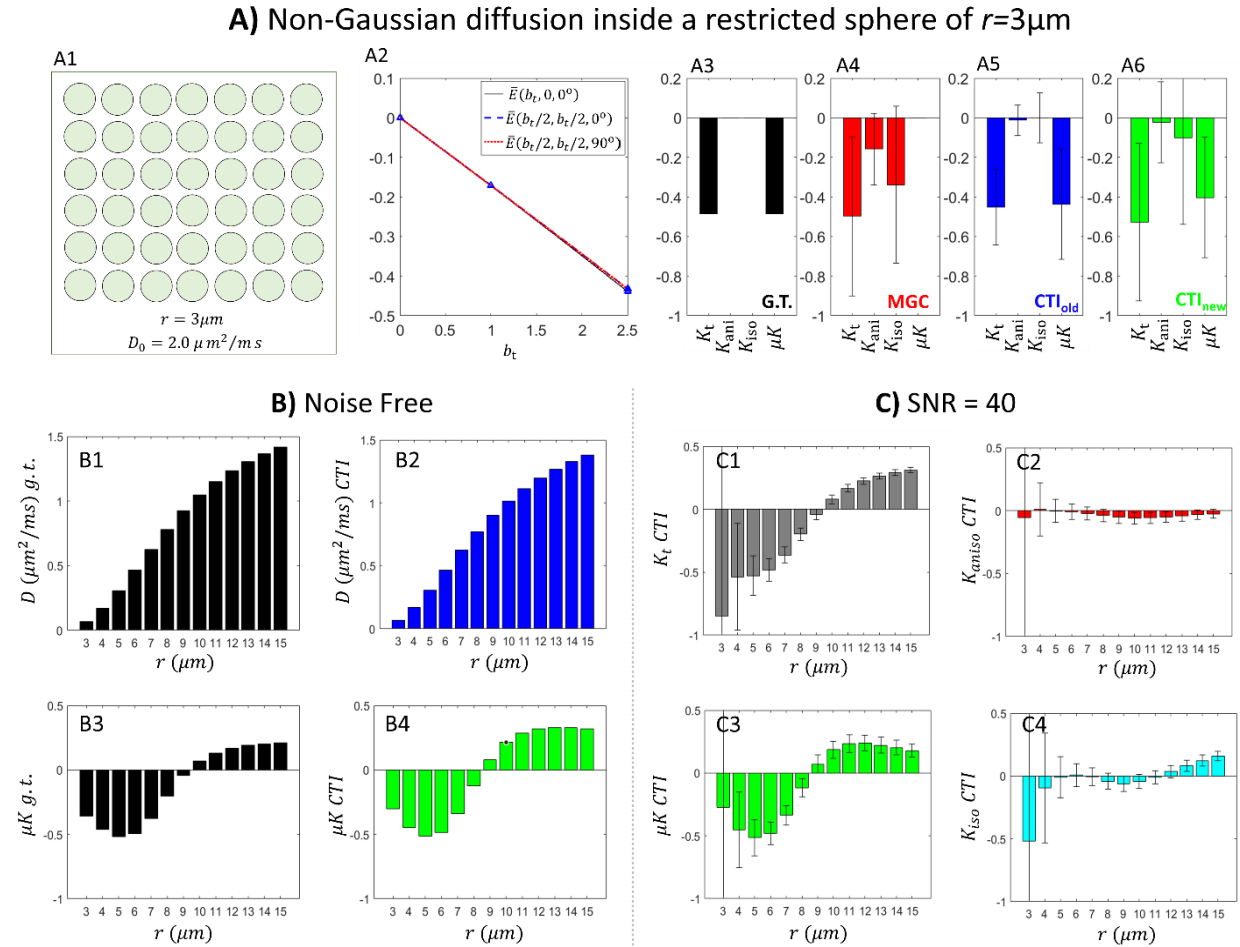

**Figure S1 – CTI estimates of a single spherical compartment. A)** Non-Gaussian diffusion inside restricted spheres with radius of $3\mu m$ and intrinsic diffusivity of $2\mu m^2/ms$: A1) model schematic representation; A2) the signal decays for three different DDE experiment types in which signals of the improved CTI protocol are marked by the blue triangles; A3) the kurtosis ground truth values; A4) the kurtosis estimates obtained from MGC; A5) the kurtosis estimates obtained from the CTI using its previous "old" protocol; A6) the kurtosis estimates obtained from the CTI using its improved "new" protocol. **B)** Results from noise free synthetic signals as a function of $r$: B1) ground truth mean diffusivity; B2) CTI measured mean diffusivity; B3) ground truth microscopic kurtosis; B4) CTI measured microscopic kurtosis. **C)** Results from noise corrupted signals (SNR=40, mean and std values from 1000 iterations) as a function of r: C1) CTI total kurtosis $K_t$; C2) CTI anisotropic kurtosis $K_{aniso}$; C3) CTI microscopic kurtosis $\mu K$; C4) CTI measured isotropic kurtosis $K_{iso}$.

To assess the robustness of the estimates to noise, CTI is also applied to the geometrical signals corrupted by Rician noise (SNR=40). Mean and standard deviations of each kurtosis estimate were computed based on 1000 simulations repetitions. Results of the noise corrupted simulations are shown in panel B of Supporting Information Figure S1 for CTI total kurtosis (C1), anisotropic kurtosis (C2), microscopic kurtosis (C3), and isotropic kurtosis (C4). Mean microscopic kurtosis estimates (B3) show analogous radius dependency than its ground truth (C3) and noise free values (C4). Lower precision in kurtosis estimates is, however, observed for lower radius which can be explained by higher estimate uncertainties of systems comprising lower diffusivities (c.f. supporting information, section D). As expected, mean isotropic and anisotropic kurtosis shows values close to zero (panels C2 and C4) – small deviations from zero at higher radius can be a consequence of higher-order terms biases, Rican noise biases, or incomplete Z tensor convergence to the covariance tensor (due to the violation of long mixing time regime).

# Section C – Considerations for the "old" CTI approach

In our previous work (Henriques et al., 2020), kurtosis source estimates were derived from a CTI acquisition comprising six pairs of $b_1$ and $b_2$ combinations, repeated for 2 sets of parallel/perpendicular directions. These correspond to the 12 experiments summarized in Supporting Information Table S1.

**Table S1** – Summary of the complete DDE parameter combination used for the CTI estimates in Henriques et al. (2020). The 12 parallel directions of the 5-design and the 60 perpendicular directions of Jespersen's scheme are reported by (Jespersen et al., 2013).

*Complete non-optimized CTI protocol*

| sets | $b_1$ | $b_2$ | $b_t$ | $\theta$ | $b_\Delta$ | direction scheme |
|---|---|---|---|---|---|---|
| #1 | $b_a$ | $b_a$ | $2b_a$ | 0° | 1 | Parallel directions from 5-design and 8-design (12+45 directions) |
| #2 | $b_a$ | $b_a$ | $2b_a$ | 90° | -1/2 | Perpendicular directions of Jespersen's DDE perpendicular scheme (60 directions) |
| #3 | $b_a$ | 0 | $b_a$ | NA | 1 | Directions from 5-design and 8-design (12+45 directions) |
| #4 | $b_a$ | 0 | $b_a$ | NA | 1 | 1st encoding directions of Jespersen's perpendicular directions scheme (60 directions) |
| #5 | 0 | $b_a$ | $b_a$ | NA | 1 | Directions from 5-design and 8-design (12+45 directions) |
| #6 | 0 | $b_a$ | $b_a$ | NA | 1 | 2nd encoding directions of Jespersen's perpendicular directions scheme (60 directions) |
| #7 | $b_b$ | $b_b$ | $2b_b$ | 0° | 1 | Parallel directions from 5-design and 8-design (12+45 directions) |
| #8 | $b_b$ | $b_b$ | $2b_b$ | 90° | -1/2 | Perpendicular directions of Jespersen's DDE perpendicular scheme (60 directions) |
| #9 | $b_b$ | 0 | $b_b$ | NA | 1 | Directions from 5-design and 8-design (12+45 directions) |
| #10 | $b_b$ | 0 | $b_b$ | NA | 1 | 1st encoding directions of Jespersen's perpendicular directions scheme (60 directions) |
| #11 | 0 | $b_b$ | $b_b$ | NA | 1 | Directions from 5-design and 8-design (12+45 directions) |
| #12 | 0 | $b_b$ | $b_b$ | NA | 1 | 2nd encoding directions of Jespersen's perpendicular directions scheme (60 directions) |

It is important to note that in Henriques et al. (2020), sets #1, #2, #7 and #8 were also repeated for $n_2$ inverted directions to evaluate the long mixing time regime approximations. These inverted directions acquisitions are excluded in Supporting Information Table S1 for the sake of simplicity.

Note that this non-optimised protocol contains redundant sets. Particularly, sets #3-6 and sets #9-12 are related to identical powder-averaged signals, i.e. $\bar{E}_{DDE}(b_a, 0) = \bar{E}_{DDE}(0, b_a)$ and $\bar{E}_{DDE}(b_b, 0) = \bar{E}_{DDE}(0, b_b)$ respectively. For the reference non-optimized CTI protocol of our current study, these redundant sets are removed, and the direction schemes of each set are adapted to ensure homoscedasticity (Supporting Information Table S2).

**Table S2** – Summary of the reference non-optimized CTI protocol used on the current study.

*Complete non-optimized CTI protocol*

| sets | $b_1$ | $b_2$ | $b_t$ | $\theta$ | $b_\Delta$ | direction scheme |
|---|---|---|---|---|---|---|
| #1 | $b_a$ | 0 | $b_a$ | NA | 1 | 45 directions of the 8-design (x3 repetitions) |
| #2 | $b_a$ | $b_a$ | $2b_a$ | 0° | 1 | 45 directions of the 8-design (x3 repetitions) for both diffusion encodings |
| #3 | $b_a$ | $b_a$ | $2b_a$ | 90° | -1/2 | 45 directions of the 8-design for the 1st encoding, repeated for 3 orthogonal directions for the 2nd encoding |
| #4 | $b_b$ | 0 | 1 | NA | 1 | 45 directions of the 8-design (x3 repetitions) |
| #5 | $b_b$ | $b_b$ | $2b_a$ | 0° | 1 | 45 directions of the 8-design (x3 repetitions) for both diffusion encodings |
| #6 | $b_b$ | $b_b$ | $2b_b$ | 90° | -1/2 | 45 directions of the 8-design for the 1st encoding, repeated for 3 orthogonal directions for the 2nd encoding |

In contract to the improved protocol of the current study, one can note that the non-optimized protocol presents set experiments with unbalanced total b-values, i.e. its symmetric DDE experiments (i.e. sets #2-3 and #5-6) are performed for a higher total b-value than SDE-equivalent experiments (i.e. sets #1 and #3).

To remain consistent with the improved protocol designed on our study, parameters $b_a$ and $b_b$ in Supporting Information Table S2 were set to $2.5 ms/\mu m^2$ and $1 ms/\mu m^2$ for the non-optimized protocol, which this leaded to a non-optimized protocol with a high total b-value of $5 ms/\mu m^2$ (sets #2-3). In our previous work, we showed that higher b-values are associated with higher-order-term effects. However, here, we show that, rather of being only an effect of the higher b-values used, the biases in $\mu K$ observed on our previous study is also a consequence of the unbalanced b-values used in the non-optimized protocol. Below, Supporting Information Figure S2 shows the simulation results for the non-optimized protocol for models 2 and 5 (models which showed to be problematic for the non-optimized protocol) applied with lower total b-values (i.e. $b_a = 1.25 ms/\mu m^2$ and $b_b = 0.5 ms/\mu m^2$), which still shows overestimation of $\mu K$, thereby supporting the notion that unbalanced b-values contribute to the biases in $\mu K$ estimation in the old CTI approach.

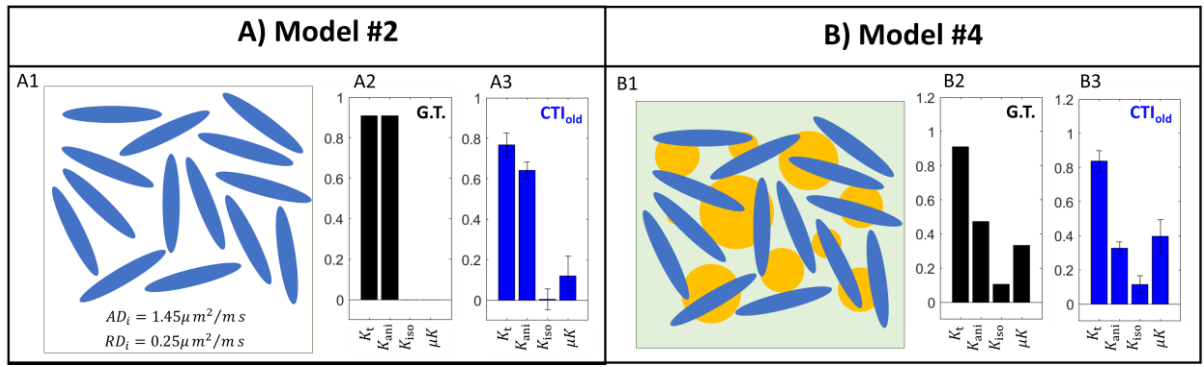

**Figure S2 – Kurtosis estimates of non-optimized CTI protocol with parameters $b_a = 1.25 ms/\mu m^2$ and $b_b = 0.5 ms/\mu m^2$ for two simulations (model 2 and 4 described in main manuscript)**; **A)** Evenly oriented anisotropic Gaussian diffusion components; **B)** Sum of different compartment types (isotropic Gaussian + anisotropic Gaussian + isotropic non-Gaussian components). From left to right, each panel shows: a schematic representation of the toy models (A1, B1); the kurtosis ground truth values (A2, B2); and the kurtosis estimates obtained from the CTI using its "old" non-optimized protocol (A3, B3).

## Section D - Precision of Microscopic Kurtosis estimates

Considering the propagation of uncertainty, the expected error of $\mu K$ ($\sigma_{\mu K}$) can be calculated from Eq. 6, which yields:

$$\sigma_{\mu K} = \frac{12}{b_a^2 D^2} \sqrt{\frac{\sigma^2}{S_1^2 N} + \frac{\sigma^2}{S_2^2 N}} \quad (S1)$$

where $S_1 = \bar{E}_{DDE}(b_a, 0)$, $S_2 = \bar{E}_{DDE}\left(\frac{b_a}{2}, \frac{b_a}{2}, 0°\right)$, $\sigma$ is the signal noise standard deviation, and $N$ is the number of pairs of directions used to compute the powder averages (here we assumed that the error of the mean diffusivity $D$ is negligible). From equation S1 one can note that $\sigma_{\mu K}$ does not only depend on acquisition parameters (b-value, number gradient direction pair) but on the ground truth diffusion parameters of the system (particularly, its mean diffusivity).

To illustrate the dependency of the $\mu K$ uncertainty with the diffusion parameters of the system, Supporting Information Figure S3 panel A and B we shows the $\sigma_{\mu K}$ values computed for different ground truth mean diffusivities $D$ and microscopic kurtosis $\mu K$ based on the acquisition parameters of this study (i.e. $N=135$ and $b_a = 2.5 ms/\mu m^2$) and for two different SNRs (40 and 20) - note, for this figure, ground truth $S_1$ and $S_2$ signals are approximated as $\bar{E}_{DDE}(b_1, b_2, 0°) = \exp\left(-(b_1 + b_2)D + \frac{1}{6}(b_1^2 + b_2^2)D^2 \mu K\right)$. These panels show that systems with lower mean diffusivities and microscopic kurtosis values are associated to lower $\mu K$ uncertainties. For the mean diffusivity and microscopic kurtosis values observed on the white matter and grey matter ROIs of our study (marked in by the red and green points respectively), predicted $\sigma_{\mu K}$ are between 0.05 and 0.1 for a SNR=40 (Supporting Information Figure S3, panel A) and between 0.1 and 0.2 for a SNR=20 (Supporting Information Figure S3, panel A) – these uncertainly levels are in line with the standard deviation on panel A4 of Figure 7.

Panel C of Supporting Information Figure S3 shows the minimal SNR requirement to obtain a $\sigma_{\mu K}=0.05$, showing the high SNR requirements to estimate microscopic kurtosis for systems of low diffusivities. These SNR requirement maps was computed using the following expression for $N=135$ and $b_a = 2.5 ms/\mu m^2$:

$$SNR = \frac{1}{\sigma_{\mu K}} \frac{12}{b_a^2 D^2} \sqrt{\frac{1}{S_1^2 N} + \frac{1}{S_2^2 N}} \quad (S2)$$

Note that Supporting Information Figure S3 can be easily adapted to other acquisition parameters using equations S1 and S2.

To support that the propagation of uncertainty analysis is consistent to the numerical simulations, in Supporting Information Figure S3 panel D we show the histogram of $\mu K$ estimates obtained from the main manuscript noise corrupted numerical simulations (signals from model 3 corrupted with Rician noise at a nominal SNR of 40) for three ground truth set of parameters representing: 1) a reference simulation with zero $\mu K$ ($D_{gt} = 0.8 \mu m^2/ms$, and $\mu K_{gt} = 0$); 2) the mean values measured for GM ($D_{gt} = 0.76 \mu m^2/ms$, and $\mu K_{gt} = 0.45$); 3) the mean values measured for WM ($D_{gt} = 0.82 \mu m^2/ms$, and $\mu K_{gt} = 0.27$). For these three sets of ground truth parameters, the standard deviation of the values reported in the histograms (0.067±0.002, 0.059±0.002, and 0.061±0.002) are consistent to the $\sigma_{\mu K}$ values obtained for the propagation of uncertainty analysis (0.068, 0.055, 0.59). It is important to note that from our numerical simulations and propagation of uncertainty analysis, noise does introduce significant offset between the $\mu K$ mean values and ground truth values and thus the measured positive $\mu K$ values measured in our study cannot be explained as being a noise artifact.

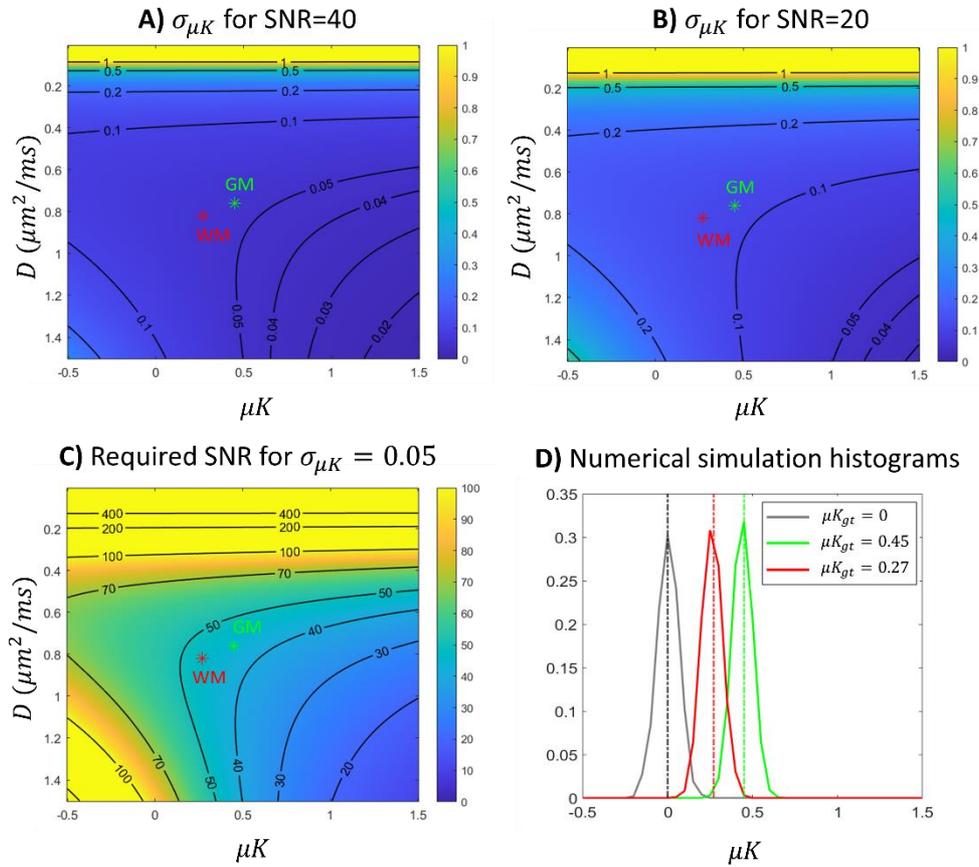

**Figure S3 – Microscopic kurtosis estimation uncertainty.** **A)** Microscopic kurtosis uncertainty $\sigma_{\mu K}$ prediction for different ground truth mean diffusivities $D$ and microscopic kurtosis $\mu K$ values and given the acquisition parameters of this study ($N$=135 and $b_a = 2.5 ms/\mu m^2$) and a single acquisition nominal SNR of 40 (i.e. $\sigma = 0.025$); **B)** $\sigma_{\mu K}$ prediction for different ground truth $D$ and $\mu K$ values and given the acquisition parameters of this study ($N$=135 and $b_a = 2.5 ms/\mu m^2$) and a single acquisition nominal SNR of 20 (i.e. $\sigma = 0.05$); **C)** Required SNR to obtain a $\sigma_{\mu K}$=0.05 and given the acquisition parameters of this study ($N$=135 and $b_a = 2.5 ms/\mu m^2$). In all panels, the mean D and $\mu K$ values observed on the white matter and grey matter ROIs of the rat brain data is marked in by the red and green points, respectively. **D)** histogram of $\mu K$ estimates obtained from signals computed using numerical simulations (signals from model 3) and corrupted with Rician noise at a nominal SNR of 40 for three ground truth set of parameters representing: 1) $D_{gt} = 0.8 \mu m^2/ms$, and $\mu K_{gt} = 0$ (histogram marked by the grey line); 2) $D_{gt} = 0.76 \mu m^2/ms$, and $\mu K_{gt} = 0.45$ (histogram marked by the green line); 3) $D_{gt} = 0.82 \mu m^2/ms$, and $\mu K_{gt} = 0.27$ (histograms marked by the red line). These latter histograms were produced for 1000 simulation iterations and the ground truth values are marked by the dashed lines.

# Section E – Auxiliary T2-weighted experiments (Methods)

The T2 weighted images in this study were acquired with the following parameters: TR = 2000 ms, effective TE = 36 ms, RARE factor = 8, Field of View = 24 × 16.1 mm$^2$, matrix size 160 × 107, leading to an in-plane voxel resolution = 150 × 150 μm$^2$, slice thickness = 500 μm (21 slices), number of averages = 8. Representative raw sagittal T2-weighted images for all three rats are shown in Supporting Information Figure S4. These images were used as a reference for placing the 3 coronal slices for diffusion MRI acquisition – positions of the coronal slices are marked in red.

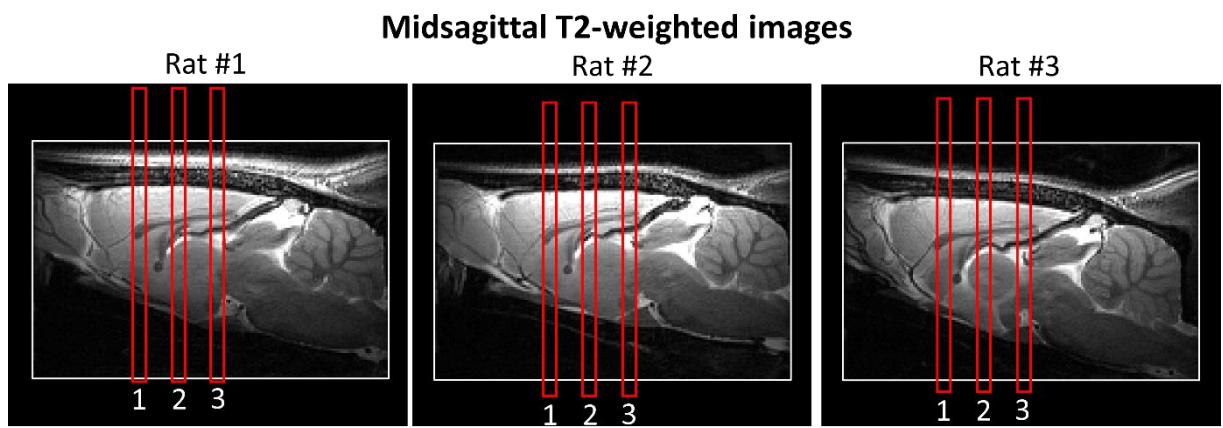

**Figure S4 – Midsagittal T2-weighted images of all three animals.** These images were used as a reference for placing the 3 coronal slices for diffusion MRI acquisition – positions of the coronal slices are marked in red.

## Section F – Long mixing time regime

To empirically check the long mixing time regime, the following two extra DDE set experiments were performed for Rat #1:

- (set #5) acquisitions for $b_1 = b_2 = 1 ms/\mu m^2$ and parallel DDE directions ($S_5 = \bar{E}_{DDE}(b_1, b_2, 0º)$). For powder-averaging 45 directions of the 8-design repeated 3 times were acquired.
- (set #6) acquisitions for $b_1 = b_2 = 1 ms/\mu m^2$ and antiparallel DDE directions ($S_6 = \bar{E}_{DDE}(b_1, b_2, 180º)$). For powder-averaging 45 directions of the 8-design repeated 3 times ware acquired (note that these directions are inverted for the 2$^{nd}$ encoding).

Other acquisition parameters were equal to the main experiments of these study. Each set was acquired with together with 24 $b_t = 0$ acquisitions for signal decay normalization. Set #5 was also fully repeated 2 times ($S_5^{\#1}$ and $S_5^{\#2}$). Supporting Information Figure S5 shows the powder-averaged signal decays for the two repetitions of set #5 (panels A and B) and for set #6 (panel C). All three experiments show identical signal decays.

The log mixing time regime can be confirmed by checking if the log differences between the powder-averaged signals of set #5 and #6 is near zero (e.g. Jespersen and Buhl, J Magn Reson 2011, Henriques et al., 2020). The maps of the log differences between the two repetitions of data acquisition for set #5 is shown in panel D (for a reference), and the maps of the log differences between signals of set #5 and #6 are shown in panel E. While B0 inhomogeneity artefacts are present in regions outside of the brain for both panels D and E (red arrows), log difference values inside the brain mask are centred in zero. As indicated by the histograms in panel F, ranges of the values for the log difference between signals of set #5 and #6 are identical to the ranges of the values for the log difference between repeated signals of set #5 acquisitions – this shows that the signal fluctuations on the log differences between parallel and antiparallel DDE acquisitions are in the same range than the log differences between repeated noisy signals.

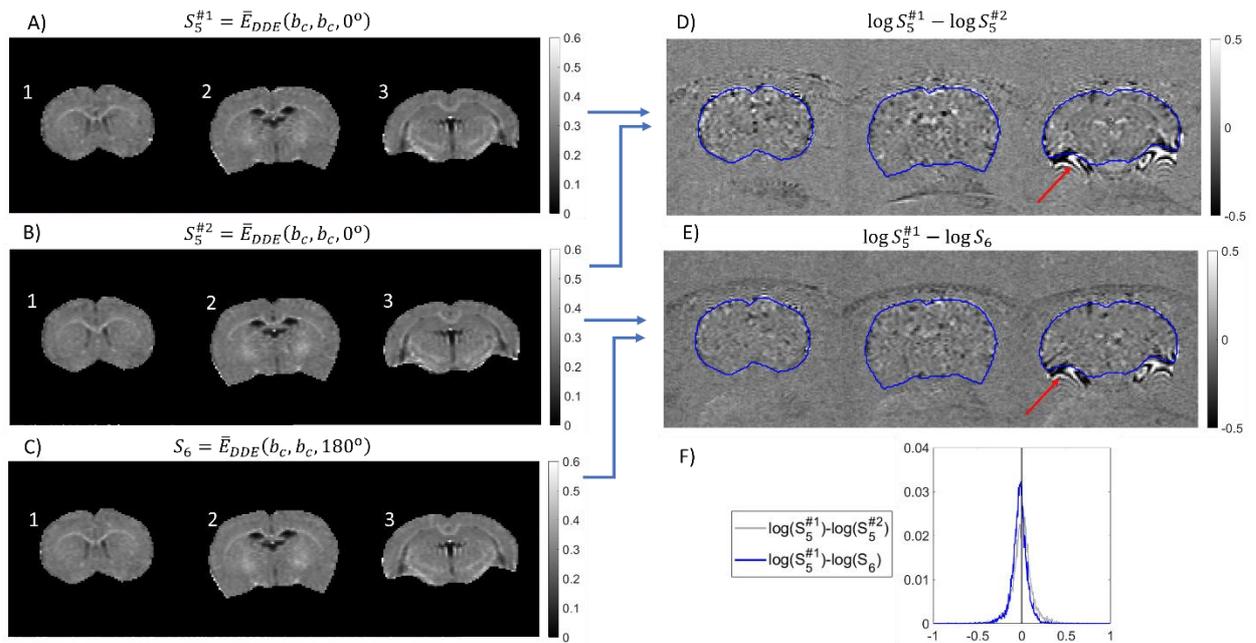

**Figure S5 – Powder-averaged data for the extra experiments performed on Rat #1. A)** Powder-averaged signal decays for DDE parallel acquisitions (repetition #1); **B)** Powder-averaged signal decays for DDE parallel acquisitions (repetition #2); **C)** Powder-averaged signal decays for DDE antiparallel acquisitions. **D)** Map of the log differences values between powder-averaged signals of set #5 repetitions #1 and #2 (brain mask region delineated in blue, and B0 inhomogeneities artifacts are pointed by the red arrow). **E)** Map of the log differences values between powder-averaged signals of set #5 and #6 brain mask region delineated in blue, and B0 inhomogeneities artifacts are pointed by the red arrow). **F)** Histograms of the values inside the brain mask for the log differences values between powder-averaged signals of set #5 repetitions #1 and #2 (grey) and the log differences between powder-averaged signals of set #5 and #6 (blue). Note parameters $b_c$ was set to $1\ ms/\mu m^2$ (i.e total b-value = $2\ ms/\mu m^2$).

## Section G – Data processing (Methods)

Thermal noise of each diffusion-weighted data set and for each cryocoil channel was suppressed using a threshold-based Marchenko-Pastur PCA denoising (92) in which signal components for eigenvalues $\lambda < \sigma^2\left(1 + \sqrt{N/M}\right)^2$ were removed - $\sigma^2$ is noise variance (computed from the repeated $b_t = 0$ acquisitions), $N$ and $M$ are the number of gradient directions pairs and image voxels. The denoised data was subsequently corrected for Gibbs ringing using a sub-voxel shift algorithm (29,93). The processed diffusion-weighted signals for the four channels were then combined using sum-of-squares. Combined data for different gradient direction pairs were then aligned along the different b-values and directions using a sub-pixel registration technique (94).

## Section H – MGC kurtosis estimates using sets #2-3

Here we report the results for MCG kurtosis estimates obtained by fitting Eq. 9 to only DDE sets #2, #3, and #4. Under this condition, one can observe that $K_{aniso}^{MGC}$ and $K_{aniso}^{CTI}$ are identical (Supporting Information Figure S6); however, $K_{iso}^{MGC}$ is still a combination of isotropic and microscopic kurtosis effects. Note that these observations can be theoretically derived from Eqs. 6 and 9 as shown below.

**MGC $K_{aniso}$**: From Eq. 9, one can note that $K_{aniso}^{MGC}$ can be resolved by:

$$\log \bar{E}_{MGC}(b, 1) - \log \bar{E}_{MGC}\left(b, -\frac{1}{2}\right) = \frac{1}{2} b^2 D^2 K_{aniso}^{MGC} \qquad (S3)$$

For $\bar{E}_{MGC}(b, 1) = \bar{E}_{DDE}\left(\frac{b}{2}, \frac{b}{2}, 0°\right)$ and $\bar{E}_{MGC}(b, 1) = \bar{E}_{DDE}\left(\frac{b}{2}, \frac{b}{2}, 90°\right)$, equation S3 is equivalent to Eq. 8 – proving that $K_{aniso}^{MGC} = K_{aniso}^{CTI}$ when MGC kurtosis estimates are obtained from data acquired with identical waveform profiles (i.e. DDE experiments sets #2-3). Note that this would not be the case if data from set #1 is used (i.e., by considering $\bar{E}_{MGC}(b, 1) = \bar{E}_{DDE}(b, 0, 0°)$).

**MGC $K_{iso}$**: $K_{iso}^{MGC}$ can be extracted by subtracting $K_{aniso}^{MGC} = K_{aniso}^{CTI}$ from the total apparent kurtosis $K_t^{app}$ of the b-value dependency of $\bar{E}_{MGC}(b, 1)$. Considering that, from the signal decays measured by set #2 and #4, $K_t^{app}$ is equal to $K_{aniso}^{cti} + K_{iso}^{cti} + \frac{\mu K}{2}$ (c.f. Eq. 6), one can show that $K_{iso}^{MGC} = K_{iso}^{cti} + \frac{\mu K}{2}$ - this proves that $K_{iso}^{MGC}$ depends on both sources of isotropic and microscopic kurtosis.

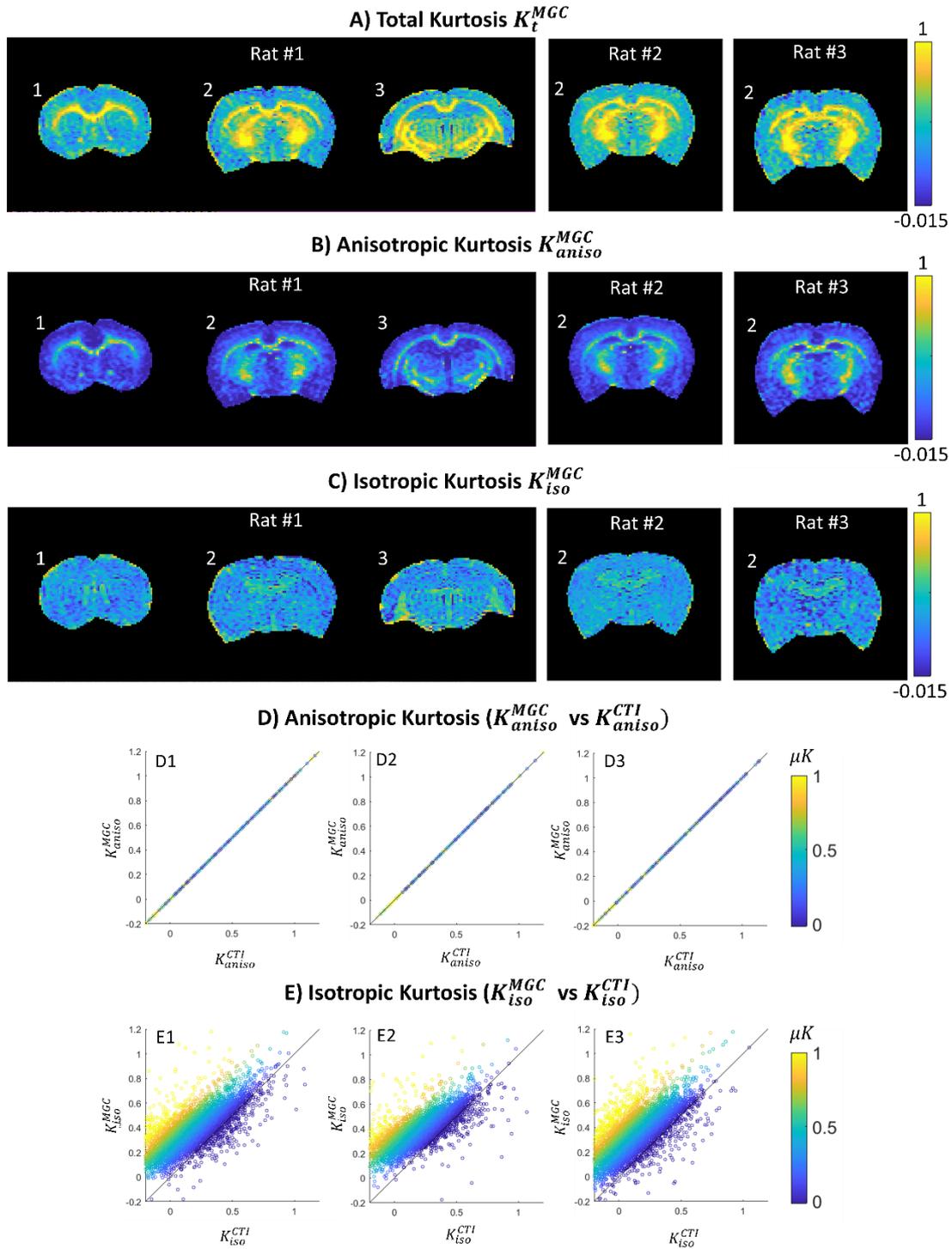

**Figure S6 – MGC kurtosis estimates using only data from sets #2, #3, #4 and its comparison to CTI estimates using all data. A)** MGC total kurtosis ($K_t^{MGC}$) maps. **B)** MGC anisotropic kurtosis $K_{aniso}$ maps. **C)** MGC isotropic kurtosis $K_{iso}$ maps. **D)** Scatter plots between MGC anisotropic kurtosis ($K_{aniso}^{MGC}$) and CTI anisotropic kurtosis ($K_{aniso}^{CTI}$) estimates; **E)** Scatter plots between MGC isotropic kurtosis ($K_{iso}^{MGC}$) and CTI isotropic kurtosis ($K_{iso}^{CTI}$) estimates.